\documentclass[12pt]{article}
\usepackage{etex}

\usepackage[utf8]{inputenc}
\usepackage[TS1,T1]{fontenc}
\usepackage{xspace}
\usepackage{amssymb,amsmath,mathtools}
\usepackage[all]{xy}
\usepackage{graphicx} 
\usepackage[a4paper,hmargin=3cm,vmargin=3cm]{geometry}
\usepackage{array,booktabs}
\usepackage{slashed}
\usepackage[english]{babel}
\usepackage{hyperref}
\usepackage{lmodern}

\usepackage{amsmath}

\usepackage{enumitem}

\setlist[itemize]{parsep=0pt,partopsep=0pt,topsep=0pt,itemsep=0pt}
\setlist[enumerate]{parsep=0pt,partopsep=0pt,topsep=0pt,itemsep=0pt}

\DeclareFontFamily{U}{matha}{\hyphenchar\font45}
\DeclareFontShape{U}{matha}{m}{n}{
      <5> <6> <7> <8> <9> <10> gen * matha
      <10.95> matha10 <12> <14.4> <17.28> <20.74> <24.88> matha12
      }{}
\DeclareSymbolFont{matha}{U}{matha}{m}{n}
\DeclareFontSubstitution{U}{matha}{m}{n}

\DeclareFontFamily{U}{mathb}{\hyphenchar\font45}
\DeclareFontShape{U}{mathb}{m}{n}{
      <5> <6> <7> <8> <9> <10> gen * mathb
      <10.95> mathb10 <12> <14.4> <17.28> <20.74> <24.88> mathb12
      }{}
\DeclareSymbolFont{mathb}{U}{mathb}{m}{n}
\DeclareFontSubstitution{U}{mathb}{m}{n}

\DeclareMathSymbol{\varstar}       {2}{matha}{"0F}
\DeclareMathSymbol{\convolution}   {2}{mathb}{"0A}

\newcommand{\spmatrix}[1]{\left( \begin{smallmatrix}#1\end{smallmatrix}\right)} % petites matrices
\newcommand{\omi}[1]{\buildrel { \buildrel{#1}\over{\vee} } \over .}

\newcommand\bbbone{{\mathchoice{1\mskip-4mu\mathrm{l}}{1\mskip-4mu\mathrm{l}}{1\mskip-4.5mu\mathrm{l}} {1\mskip-5mu\mathrm{l}}}}

\newcommand{\grast}{\bullet}

\newcommand{\invast}{\varstar}% si mathabx n'est pas installé, remplacer \varstar par \ast

\newcommand{\dualast}{\ast}

\newcommand{\hodgeast}{{\mathord{\star}}}

\newcommand{\moyast}{{\mathord{\convolution}}}% si mathabx n'est pas installé, remplacer \varstar par \star

\newcommand\adrep{{\text{\textup{ad}}}} % ad representation, \ad pris pas jpconf
\newcommand\dd{\text{\textup{d}}} % differentielles
\newcommand\hd{\widehat{\dd}} % differentielles
\newcommand\hR{\widehat{R}}
\newcommand\cdotaction{\mathord{\cdot}} % le point centré pour une action (et non un opérateur)
\newcommand\exter{{\textstyle\bigwedge}} % signe wedge en grand, pour les algèbres
\newcommand\sign{{\text{\textup{sign}}}}
\newcommand\der{{\text{\textup{Der}}}} % Derivations
\newcommand\Int{{\text{\textup{Int}}}} % Interior derivations
\newcommand\Out{{\text{\textup{Out}}}}

\DeclareMathOperator{\Aut}{Aut} % automorphismes
\DeclareMathOperator{\Hom}{\mathsf{Hom}} % foncteur Hom
\DeclareMathOperator{\tr}{Tr} % trace

\newcommand\varnotation[1]{{\mathcal{#1}}}
\newcommand\algnotation[1]{{\mathbf{#1}}}
\newcommand\grnotation[1]{{\mathsf{#1}}}
\newcommand\modnotation[1]{{\boldsymbol{#1}}}
\newcommand\ehnotation[1]{{\mathcal{#1}}}

\newcommand\varE{{\varnotation{E}}}
\newcommand\varM{{\varnotation{M}}}
\newcommand\varP{{\varnotation{P}}}
\newcommand\algA{{\algnotation{A}}}
\newcommand\algzero{{\grnotation{0}}}
\newcommand\modM{{\modnotation{M}}}
\newcommand\ehH{{\ehnotation{H}}}
\newcommand\ka{{\mathfrak a}}
\newcommand\kb{{\mathfrak b}}
\newcommand\kg{{\mathfrak g}}
\newcommand\kS{{\mathfrak S}} 
\newcommand\kX{{\mathfrak X}}
\newcommand\kY{{\mathfrak Y}}
\newcommand\ksl{{\mathfrak{sl}}}
\newcommand\ksu{{\mathfrak{su}}}
\newcommand\kisp{{\mathfrak{isp}}}
\newcommand\gH{{\mathbb H}}
\newcommand\gN{{\mathbb N}}
\newcommand\gR{{\mathbb R}}
\newcommand\gC{{\mathbb C}}
\newcommand\caA{{\mathcal A}}
\newcommand\caB{{\mathcal B}}
\newcommand\caD{{\mathcal D}}
\newcommand\caG{{\mathcal G}}
\newcommand\caH{{\mathcal H}}
\newcommand\caL{{\mathcal L}}
\newcommand\caR{{\mathcal R}}
\newcommand\caS{{\mathcal S}}
\newcommand\caU{{\mathcal U}}
\newcommand\caZ{{\mathcal Z}}

\usepackage{amsthm,thmtools}

\theoremstyle{definition}

\theoremstyle{remark}

\usepackage[square,numbers,sort,compress,semicolon,merge]{natbib}
\hypersetup{
pdfauthor={Thierry Masson},
pdftitle={Gauge theories in noncommutative geometry},
pdfsubject={gauge theory, noncommutative geometry},
pdfcreator={pdflatex},
pdfproducer={pdflatex},
pdfkeywords={gauge theory, noncommutative geometry, differential calculus, spectral triple, derivations}
}

\hypersetup{
plainpages=false,
colorlinks=true,% à mettre avant la suite : supprime l'encadrement des liens
linkcolor=black, 
anchorcolor=black, 
citecolor=black, 
urlcolor=black, 
menucolor=black, 
filecolor=black, 
bookmarksopen=true,
bookmarksnumbered=true}

\begin{document}

%%%%%%%%%%%%%%%%%%%%%%%%%%
% standard LaTeX
\title{Gauge theories in noncommutative geometry}
\author{T. Masson$^a$}
\date{\today}
\maketitle
\begin{center}
$^a$ Centre de Physique Th\'eorique\\
Aix-Marseille Univ, CNRS UMR 7332, Univ Sud Toulon Var\\
Case 907 - Campus de Luminy\\
F-13288 Marseille Cedex 9
\end{center}
%%%%%%%%%%%%%%%%%%%%%%%%%%

\bigskip
\begin{abstract}
In this review we present some of the fundamental mathematical structures which permit to define noncommutative gauge field theories. In particular, we emphasize the theory of noncommutative connections, with the notions of curvatures and gauge transformations. Two different approaches to noncommutative geometry are covered: the one based on derivations and the one based on spectral triples. Examples of noncommutative gauge field theories are given to illustrate the constructions and to display some of the common features.
\end{abstract}

\maketitle

%\setcounter{tocdepth}{1}
%\tableofcontents

%%%%%%%%%%%%%%%%%%%%%%%%%%%%%%%%%%%%%%%%%%%%%%%%%%%%%%%%%%%%%%%%%%%%%%%%%%
\section{Introduction}
%\label{}
%%%%%%%%%%%%%%%%%%%%%%%%%%%%%%%%%%%%%%%%%%%%%%%%%%%%%%%%%%%%%%%%%%%%%%%%%%

Local gauge symmetries are an essential ingredient of model building techniques in today high energy physics. The gauge invariance principle determines in an economical way the structure of the fundamental interactions modeled in the Standard Model of particle physics. 

Yang-Mills theories have been recognized to be mathematically the theory of connections of principal fiber bundles. This identification is part of the global understanding of the mathematical structures used in the Standard Model of particles physics, like spinors and Dirac operators on the matter side.

Since its emergence in the 80's \cite{Conn85, Connes2000by, ConnMarc08a, DuVi01a, DuVi91, Land97a, GracVariFigu01}, noncommutative geometry has helped to reveal deep mathematical relationships between ordinary geometry and other structures, among them differential algebras and normed algebras. In particular, noncommutative geometry has shed new lights on gauge theories. Indeed, a theory of connections can be defined in great generality using the noncommutative language of associative algebra, modules and differential calculi.

Different approaches have been proposed to study noncommutative spaces. The theory of spectral triples, developed by Connes, emphasizes the metric structure \cite{Conn94, ConnMarc08a, GracVariFigu01}. On the other hand, many noncommutative spaces are studied through differential structures \cite{DuVi01a, DuVi88, MR1103056, DuVi91, MR1205605, MR1150534, Mass30, Mass33}. However, all the noncommutative gauge field theories studied so far use the same building blocks, even when there are defined through different approaches.

Moreover, many of these gauge theories, independently of their exact constitutive elements, share some common or similar features. One of them is the origin of the gauge group. Another one, and not the least, is the possibility to naturally produce Yang-Mills-Higgs Lagrangians.

In this review, we focus on the structures behind noncommutative gauge field theories: the notion of connections, the identification of the gauge group and the definition of an invariant action. Representative examples of different approaches are presented.

Before diving into the noncommutative world, it is worth recalling the main features of gauge field theories and the main mathematical structures used to model them. See \cite{o2000gauge} for a historical review.

A gauge interaction is an implementation of the principle that the theory should be invariant under some \emph{local} symmetry. In particle physics, these local symmetries take the form of functions $g : \varM \rightarrow G$ on the space-time $\varM$ with values in a structure group $G$. Electromagnetism is associated to the group $G=U(1)$, the electroweak theory by Glashow, Weinberg and Salam uses the group $G=U(1) \times SU(2)$ and chromodynamics relies on $G=SU(3)$ \cite{PeskSchr08a}. 

Asking for invariance under a global symmetry in a group $G$ is done using some invariants in the representation theory of (here mainly compact) groups. But the fulfillment of the local version of the theory requires the introduction of some auxiliary vector fields: the gauge potentials $A_\mu$. Under a gauge transformation induced by $g$, these new fields take care in the Lagrangian for the extra terms coming from the derivatives of the non constant group elements $g$. These gauge fields are put in the Lagrangian at very specific places through the so-called minimum coupling, which consists to replace all the partial derivatives $\partial_\mu$ by the covariant derivatives $\nabla_\mu = \partial_\mu + ie A_\mu$.

From a mathematical point of view, the fields $A_\mu$ are the local descriptions of a global connection $1$-form $\omega \in \Omega^1(\varP) \otimes \kg$ on a principal fiber bundle $\varP$ over $\varM$ with structure group $G$, where $\kg$ is the Lie algebra of $G$. The gauge transformations have two identifications in this framework. The first one consider them as  passive transformations: they are the transformations between different local descriptions $A_\mu$ of the global object $\omega$. In this sense, gauge transformations are just a generalization of some change of coordinates (the correct terminology is ``change of trivializations'' in this setting). The second identification considers a gauge transformation as a vertical automorphism of $\varP$. Here, the action of the gauge group is active in the sense that it moves the points of $\varP$ and it also moves the related structures, like the connection $1$-form $\omega$.

The covariant derivative $\nabla$ can be looked at as the implementation of the connection $1$-form as a small (infinitesimal) displacement in some vector bundle $\varE$ associated to $\varP$. On the space of sections of $\varE$ (matter fields), this covariant derivative is a globally defined first order differential operator. The gauge group acts on this space of sections in the active way. This action is compatible with the covariant derivative in the sense that the section $\nabla \Psi$ supports the same representation as the section $\Psi$ of $\varE$. Passive gauge transformations correspond to the relations between different local descriptions of a section $\Psi$. 

All these mathematical structures are now well understood and they give rise to the rich theory of fiber bundles with connections \cite{KobaNomi96c, Bert96}. Noncommutative geometry being an extension of differential geometry, it naturally generalizes this theory of fiber bundles and connections. What is astonishing is that this generalization is very elegant, very powerful and very effective, not only from a mathematical point of view, but also in its applications to physics.

%%%%%%%%%%%%%%%%%%%%%%%%%%%%%%%%%%%%%%%%%%%%%%%%%%%%%%%%%%%%%%%%%%%%%%%%%%
\section{Noncommutative structures}
%\label{}
%%%%%%%%%%%%%%%%%%%%%%%%%%%%%%%%%%%%%%%%%%%%%%%%%%%%%%%%%%%%%%%%%%%%%%%%%%

This section presents some of the fundamental noncommutative structures which help formulate noncommutative gauge field theories. More details can be found in \cite{Conn94, DuVi91, DuVi01a, ConnMarc08a, Land97a, GracVariFigu01, Mass28}.

%%%%%%%%%%%%%%%%%%%%%%%%%%%%%%%%%%%%%%%%%%%%%%%%%%%%%%%%%%%%%%%%%%%%%%%%%%
\subsection{Noncommutative geometry in a nutshell}
%\label{}

Noncommutative geometry is more a line of research than a theory. Schematically, noncommutative geometry proceeds in three steps.

The first one is to study measurable, topological or geometric spaces not directly at the level of points, but at the dual level of algebras of functions on these spaces. In a more abstract language, we replace a category of spaces by a dual category of (commutative) algebras. This step relies heavily on fundamental theorems which assure us that a convenient algebra of functions (with extra structures) characterizes completely the kind of space we want to study \cite{Blac06, Take02c, BratRobi02c, KadiRing97c}. For instance, the Gelfand-Naïmark theorem tells us that a unital commutative $C^\ast$-algebra is always the commutative algebra of continuous functions on a compact topological space, equipped with the sup norm. A similar result on measurable spaces tells us that these spaces can be studied as commutative von~Neumann algebras. Differentiable manifolds can be studied with the help of their Fréchet algebras of smooth functions, but there is no theorem which characterizes differentiable manifolds as commutative algebras in a good category.

The second step is to find a way to study and characterize some of the properties of the spaces using only their associated commutative algebras of functions. Again, this step relies on deep theorems in mathematics, for instance in $K$-theory \cite{RordLarsLaus00, HigsRoe04, Wegg93, Blac98} and in cyclic homology \cite{CuntKhal97, Loda98}. Notice that some of the constructions performed on spaces (quotient, fibrations, etc.) should also be redefined in a more algebraic way. Concerning the theory of connections that we will consider here, the main theorem is the one by Serre and Swan \cite{Serr55a, Swan62a} which tells us how to identify in a algebraic way a vector bundle on a topological (or smooth) compact manifold through its space of sections. In the same way, connections themselves are completely rewritten in terms of algebras, modules and differential calculi.

Finally, the last step is to revoke the assumption on the commutativity of the algebra. This step relies on the fact that many of the tools and constructions used in step two make sense also for noncommutative algebras. A ``noncommutative space'' is then a noncommutative algebra in a precise category on which we can apply some machinery to study it \textsl{as if} it was the (commutative) algebra of functions on an ordinary space.

Gauge field theories in noncommutative geometry use different kind of generalizations of manifolds so that the different approaches can look very different. But the heart of the theory is always the same. In the following, we introduce the common structures which define noncommutative connections.

%%%%%%%%%%%%%%%%%%%%%%%%%%%%%%%%%%%%%%%%%%%%%%%%%%%%%%%%%%%%%%%%%%%%%%%%%%
\subsection{Algebraic structures}
\label{subsec-Algebraicstructures}

We suppose the reader familiar with the notions of associative algebras and modules \cite{Jaco85}. We denote by $\bbbone$ the unit in a unital associative algebra.

A graded algebra is a associative algebra $\algA^\grast = \bigoplus_{n\geq 0} \algA^n$ such that $a_p b_q \in \algA^{p+q}$ for any $a_p \in \algA^p$ and $b_q \in \algA^q$. Notice then that $\algA^0$ is an associative algebra and that the vector spaces $\algA^p$ are $\algA^0$-bimodules. If $\algA^\grast$ is unital, then $\bbbone \in \algA^0$ so that $\algA^0$ is unital.

A graded commutative algebra is a graded algebra for which $a_p b_q = (-1)^{pq} b_q a_p$. The algebra $\algA^0$ is then a commutative algebra, in the ordinary sense.

A graded differential algebra $(\algA^\grast, \dd)$ is a graded algebra $\algA^\grast$ equipped with a linear map of degree $+1$, $\dd : \algA^p \rightarrow \algA^{p+1}$, such that $\dd (a_p b_q) = (\dd a_p) b_q + (-1)^{p} a_p (\dd b_q)$.

A differential calculus on an associative algebra $\algA$ is a graded differential algebra $(\Omega^\grast, \dd)$ such that $\Omega^0 = \algA$. The space $\Omega^p$ is called the space of noncommutative $p$-forms (or $p$-forms in short). It is a $\algA$-bimodule.

If $\algA$ has an involution $a \mapsto a^\invast$, we suppose that the graded algebra $\Omega^\grast$ has also an involution, which we also denote by $\omega_p \mapsto \omega_p^\invast$ and which satisfies $(\omega_p \eta_q)^\invast = (-1)^{pq} \eta_q^\invast \omega_p^\invast$ for any $\omega_p \in \Omega^p$ and $\eta_q \in \Omega^q$. We suppose that the differential operator $\dd$ is real for this involution: $(\dd \omega_p)^\invast = \dd (\omega_p^\invast)$.

Let $\varM$ be a smooth manifold, and let $\algA = C^\infty(M)$ be the space of smooth functions on $\varM$. Then the de~Rham complex $(\Omega^\grast(M), \dd)$ is a differential calculus on $C^\infty(M)$.

There are many ways to define a differential calculus given an associative algebra. One of the great deal of differentiable noncommutative geometry is to define a convenient differential calculus on the algebra under study. Such a differential calculus has to be adapted to the structures and to the tools used to study the corresponding noncommutative space.

For instance, the theory of quantum groups makes great use of this concept. In that framework, the derivation rule for the differential is adapted to take into account the deformation parameter which defines the quantum group (see \cite{CharPres94,KlimSchm97} and references therein).

One can associate to any unital associative algebra $\algA$ a so-called universal differential calculus. We denote it by $(\Omega^\grast_U(\algA), \dd_U)$ (see for instance \cite{MR1443921} for a concrete construction). 

In short, it is defined as the free unital graded differential algebra generated by $\algA$ in degree $0$. The unit in $\Omega^\grast_U(\algA)$ is also a unit for $\Omega^0_U(\algA) = \algA$, so that it coincides with the unit $\bbbone$ of $\algA$. 

By construction, this differential calculus has an universal property formulated as follows. For any unital differential calculus $(\Omega^\grast, \dd)$ on $\algA$, there exists a unique morphism of unital differential calculi $\phi : \Omega^\grast_U(\algA) \rightarrow \Omega^\grast$ (of degree $0$) such that $\phi(a) = a$ for any $a\in \algA = \Omega^0_U(\algA) = \Omega^0$. This universal property permits to characterize all the differential calculi on $\algA$ generated by $\algA$ in degree $0$ as quotients of the universal one. Indeed, if $\Omega^\grast$ is generated by $\algA$ then the universal map $\phi$ is surjective and $\Omega^\grast = \Omega^\grast_U(\algA)/\ker \phi$, where $\ker \phi$ is a differential ideal in $\Omega^\grast_U(\algA)$.

An explicit construction of $(\Omega^\grast_U(\algA), \dd_U)$ characterizes forms in $\Omega^n_U(\algA)$ as finite sum of elements of the form $a \dd_U b_1 \cdots \dd_U b_n$ for $a, b_1, \dots, b_n \in \algA$. Here, the notation $\dd_U b$ is more or less formal, except that we must take into account the important relation $\dd_U \bbbone = 0$ which is a consequence of the definition of the differential calculus: $\dd_U \bbbone = \dd_U (\bbbone \bbbone) = (\dd_U \bbbone) \bbbone + \bbbone (\dd_U \bbbone) = 2 \dd_U \bbbone$ implies $\dd_U \bbbone = 0$. In an abstract language, $\dd_U$ maps any element in $\algA$ into its projection in the quotient vector space $\overline{\algA} = \algA/(\gC \bbbone)$. As vector spaces, one has $\Omega_U^n(\algA) \simeq \algA \otimes \overline{\algA}^{\otimes n}$.

Even if the associative algebra $\algA$ is commutative, the graded algebra $\Omega^\grast_U(\algA)$ is never graded commutative.

If $\algA$ is involutive, one can define on $\Omega^\grast_U(\algA)$ an involution by
\begin{equation*}
(a \dd_U b_1 \cdots \dd_U b_n)^\invast = (-1)^{\tfrac{n(n-1)}{2}} (\dd_U b_n^\invast) \cdots (\dd_U b_1^\invast ) a^\invast.
\end{equation*}

This differential calculus is very useful in mathematics: it appears for instance in Hochschild and cyclic homology \cite{CuntKhal97, Loda98, Mass31}.

Let us describe the universal differential calculus $(\Omega^\grast_U(\algA), \dd_U)$ for a commutative algebra $\algA$ of functions on a space $X$. We do not require any condition on these functions, neither continuity nor smoothness. Anyway, universal forms do not see any smooth structure. The cases when $X$ is a finite space is already very instructive to get a close understanding of the space of universal forms.

We first identify $\algA \otimes \algA$ with the space of functions on $X \times X$ by $(f \otimes g) (x_1, x_2) = f(x_1)g(x_2)$ for any $f,g \in \algA$ and any $x_1, x_2 \in X$. This can be repeated to identify $\algA \otimes \cdots \otimes \algA = \algA^{\otimes n}$ with the space of functions on $X \times \cdots \times X$. Using this identification, we define a product $\algA^{\otimes (p+1)} \otimes \algA^{\otimes (q+1)} \rightarrow \algA^{\otimes (p+q+1)}$ by $(fg)(x_1, \dots, x_{p+q+1}) = f(x_1, \dots, x_{p+1})g(x_{p+1}, \dots, x_{p+q+1})$. Now we can identify $\Omega^p_U(\algA)$ as a subspace of $\algA^{\otimes (p+1)}$ of the elements $f$ such that $f(x_1, \dots, x_{i-1}, x_i, x_i, x_{i+1}, \dots, x_{p}) = 0$ for any $x_k \in X$. These functions vanish when they are evaluated on two consecutive same points.

The differential $\dd_U$ can be implemented on $f \in \Omega^p_U(\algA) \subset \algA^{\otimes (p+1)}$ as
\begin{equation*}
(\dd_U f)(x_1, \dots, x_{p+2}) = \sum_{i=1}^{p+2} (-1)^{i+1} f(x_1, \dots, x_{i-1}, x_{i+1}, \dots, x_{p+2}).
\end{equation*}
In particular, for $f \in \algA = \Omega^0_U(\algA)$, one has $(\dd_U f)(x_1, x_2) = f(x_2) - f(x_1)$. Using ths universal property of $(\Omega^\grast_U(\algA), \dd_U)$, we see that this finite difference is the prototype of many differentials. For instance, the de~Rham differential is the infinitesimal version of this finite difference.

%For functions with values in $\gC$, one can define the involution as
%\begin{equation*}
%f \mapsto f^\invast(x_1, \dots, x_{p+1}) = (-1)^p \overline{f}(x_{p+1}, \dots, x_1)
%\end{equation*}

For $X = \{p\}$, one has $\algA = \gC$ and the only non zero space of universal forms is $\Omega^0_U(\gC) = \gC$.

%%%%%%%%%%%%%%%%%%%%%%%%%%%%%%%%%%%%%%%%%%%%%%%%%%%%%%%%%%%%%%%%%%%%%%%%%%
\subsection{Noncommutative connections}
\label{subsec-Noncommutativeconnections}

The notion of noncommutative connections relies on the use of three ingredients:
\begin{enumerate}
\item An associative algebra $\algA$.
\item A differential calculus $(\Omega^\grast, \dd)$ over $\algA$.
\item A right $\algA$-module $\modM$.
\end{enumerate}

Given these data, a noncommutative connection on $\modM$ is a linear map $\widehat{\nabla} : \modM \rightarrow \modM \otimes_\algA \Omega^1$ such that $\widehat{\nabla}(ma) = (\widehat{\nabla} m) a + m \otimes \dd a$ for any $m \in \modM$ and $a \in \algA$. This map can be extended into a map $\widehat{\nabla} : \modM \otimes_\algA \Omega^p \rightarrow \modM \otimes_\algA \Omega^{p+1}$ for any $p \geq 0$ using the derivation rule $\widehat{\nabla} (m \otimes \omega_p) = (\widehat{\nabla} m) \otimes \omega_p + m \otimes \dd \omega_p$ for any $\omega_p \in \Omega^p$.

The curvature of $\widehat{\nabla}$ is defined as $\widehat{R} = \widehat{\nabla}^2 = \widehat{\nabla} \circ \widehat{\nabla} : \modM \rightarrow \modM \otimes_\algA \Omega^2$. A straightforward computation show that $\widehat{R} (ma) = (\widehat{R} m) a$ thanks to the derivation rule defining $\widehat{\nabla}$.

The space of noncommutative connections on $\modM$ is an affine space modeled on the vector space $\Hom^\algA(\modM,\modM \otimes_\algA \Omega^1)$ of right $\algA$-modules morphisms.

Let $\algA$ has an involution. Then a Hermitian structure on $\modM$ is a $\gR$-bilinear map $\langle -, - \rangle : \modM \otimes \modM \rightarrow \algA$ such that $\langle ma, nb \rangle = a^\invast \langle m, n \rangle b$ and $\langle m, n \rangle^\invast = \langle n, m \rangle$ for any $a,b \in \algA$ and $m,n \in \modM$. One can extend $\langle -, - \rangle$ to $(\modM \otimes_\algA \Omega^p) \otimes (\modM \otimes_\algA \Omega^q) \rightarrow \Omega^{p+q}$ by $\langle m \otimes \omega_p, n \otimes \eta_q \rangle = \omega_p^\invast \langle m, n \rangle \eta_q$. Then a noncommutative connection $\widehat{\nabla}$ is said to be compatible with $\langle -, - \rangle$ if, for any $m,n\in \modM$,
\begin{equation*}
\langle \widehat{\nabla} m, n \rangle + \langle m, \widehat{\nabla} n \rangle = \dd \langle m, n \rangle.
\end{equation*}

The gauge group $\caG$ of $\modM$ is defined as the group of automorphism of $\modM$ as a right $\algA$-module: $\Phi \in \caG$ satisfies $\Phi(ma) = \Phi(m)a$ for any $m \in \modM$ and $a \in \algA$. A gauge transformation $\Phi$ can be extended to a right $\Omega^\grast$-module automorphism on $\modM \otimes_\algA \Omega^\grast$ by $\Phi(m \otimes \omega) = \Phi(m) \otimes \omega$. The action of the gauge group on the space of noncommutative connections on $\modM$ is defined as $\widehat{\nabla} \mapsto \widehat{\nabla}^\Phi = \Phi^{-1} \circ \widehat{\nabla} \circ \Phi$.

A gauge transformation $\Phi$ is said to be compatible with (or preserve) a Hermitian structure $\langle -, - \rangle$ on $\modM$ if $\langle \Phi(m), \Phi(n) \rangle = \langle m, n \rangle$ for any $m,n \in \modM$. We denote by $\caU(\caG)$ the subgroup of $\caG$ of the gauge transformations which preserve $\langle -, - \rangle$.

Once a noncommutative connection is given by the preceding procedure, a gauge theory is defined with the help of a Lagrangian density and a convenient integration. This last step depends heavily on the concrete situation, as will be seen in the next two sections.

%%%%%%%%%%%%%%%%%%%%%%%%%%%%%%%%%%%%%%%%%%%%%%%%%%%%%%%%%%%%%%%%%%%%%%%%%%
\subsection{Examples}
\label{sec-examples-ncconnections}

As a first example, let us show how the ordinary theory of connections fits into this framework.

Let us consider a compact smooth manifold $\varM$. The ordinary theory of connection is defined on vector bundles over $\varM$. Let $\varE$ be such a complex finite rank vector bundle, equipped with a Hermitian metric $h$. We denote by $\Gamma(\varE)$ the space of smooth sections of $\varE$.

An ordinary connection $\nabla$ on $\varE$ associates (in a linear way) to any vector field $X \in \Gamma(T\varM)$ on $\varM$ a linear map $\nabla_X : \Gamma(\varE) \rightarrow \Gamma(\varE)$ such that $\nabla_X (f s) = (X \cdotaction f) s + f \nabla_X s$ and $\nabla_{f X} s = f \nabla_X s$ for any $f \in C^\infty(\varM)$ and $s \in \Gamma(\varE)$.

In the spirit of noncommutative geometry, we consider the commutative algebra $\algA = C^\infty(\varM)$ of smooth functions on $\varM$ and the module $\modM = \Gamma(\varE)$ of smooth sections of $\varE$. The natural differential calculus to consider here is the de~Rham differential calculus $(\Omega^\grast(\varM), \dd)$ on $\varM$, for which one has, as expected, $\Omega^0(\varM) = C^\infty(\varM) = \algA$. Then $\modM \otimes_\algA \Omega^\grast(\varM) = \Omega^\grast(\varM, \varE)$ is the space of de~Rham forms on $\varM$ with values in the vector bundle $\varE$. It is well-known that the connection $\nabla$ can be extended as a map $\nabla : \Omega^p(\varM, \varE) \rightarrow \Omega^{p+1}(\varM, \varE)$ such that $\nabla (\omega_p \eta_q) = (\nabla \omega_p) \eta_q + (-1)^p \omega_p \dd \eta_q$ for any $\omega_p \in \Omega^p(\varM, \varE)$ and $\eta_q \in \Omega^q(\varM)$. The curvature is just $R = \nabla^2 : \Omega^p(\varM, \varE) \rightarrow \Omega^{p+2}(\varM, \varE)$.

A connection $\nabla$ on $\varE$ is then a noncommutative connection for the algebra $C^\infty(\varM)$, the differential calculus $(\Omega^\grast(\varM), \dd)$, and the module $\Gamma(\varE)$.

Let us consider now the general case of an unital associative algebra $\algA$ with a differential calculus $(\Omega^\grast, \dd)$. Let us describe noncommutative connections for different modules.

First, let us consider the right $\algA$-module $\modM=\algA$. In that case, $\modM \otimes_\algA \Omega^\grast = \Omega^\grast$. A noncommutative connection $\widehat{\nabla}$ on $\algA$ is completely given by the $1$-form $\widehat{\nabla} \bbbone = \omega \in \Omega^1$: $\widehat{\nabla} a = \widehat{\nabla} (\bbbone a) = (\widehat{\nabla} \bbbone) a + \bbbone \otimes \dd a = \omega a + \bbbone \otimes \dd a$ for any $a \in \algA$. $\omega$ is called the connection $1$-form of $\widehat{\nabla}$. The curvature of this connection is the left multiplication by the $2$-form $\Omega = \dd \omega + \omega \omega \in \Omega^2$. A element $\Phi$ in the gauge group is completely given by its value on $\bbbone$, which we denote by $\Phi(\bbbone) = g \in \algA$. This is an invertible element in $\algA$. It acts on the right module $\algA$ by multiplication on the left: $\Phi(a) = g a$. The connection $1$-form associated to $\widehat{\nabla}^\Phi$ is given by $\omega^g = g^{-1} \omega g + g^{-1} \dd g$, which is the usual gauge transformation relation on the space of connection $1$-forms on a principal fiber bundle. The curvature of $\omega^g$ is given by $g^{-1}(\dd \omega + \omega \omega) g = g^{-1} \Omega g$, so that the curvature $2$-form has homogeneous transformation rules.

When $\algA$ is involutive, the module $\algA$ has a canonical Hermitian stucture given by $\langle a, b \rangle = a^\invast b$. The associated gauge subgroup is given by $\caU(\algA) = \{ u \in \algA \ / \ u^\invast u = u u^\invast = \bbbone \}$, the group of unitary elements in $\algA$.

As a second special case of right $\algA$-modules, let us consider the free right $\algA$-module $\algA^N$ for an integer $N>0$. We denote by $e_i = (0, \dots, \bbbone, \dots, 0)$, for $i=1,\dots, N$, a canonical basis of this right module. It is convenient to look at $m= e_i a^i \in \modM$ as a column vector for the $a^i$'s, so that we can use matrix notations. One has the natural identification $\algA^N \otimes_\algA \Omega^\grast = (\Omega^\grast)^N$. The differential $\dd$ extends to a map $\dd : \algA^N \rightarrow (\Omega^1)^N$. A noncommutative connection on $\algA^N$ is completely given by a $N\times N$ matrix of $1$-forms $\omega = (\omega_i^j)_{i,j} \in M_N(\Omega^1)$ defined by $\widehat{\nabla} e_i = e_j \otimes \omega_i^j$. Then one has $\widehat{\nabla} (e_i a^i) = e_j \otimes \omega_i^j a^i + e_i \otimes \dd a^i$, which can be written in matrix notations as $\widehat{\nabla} m = \dd m + \omega m$. The curvature is the multiplication on the left on $\algA^N$ by the matrix of $2$-forms $\Omega = \dd \omega + \omega \omega \in M_N(\Omega^2)$. The gauge group identifies with $GL_N(\algA)$ which acts on $\algA^N$ by matrix multiplication (on the left). The action of $g \in GL_N(\algA)$ on the connection and on the curvature is given by $\omega^g = g^{-1} \omega g + g^{-1} \dd g$ and $\Omega^g = g^{-1} \Omega g$.

When $\algA$ is involutive, the module $\algA^N$ has a canonical Hermitian stucture given by $\langle (a^i), (b^j) \rangle = \sum_{i=1}^N (a^i)^\invast b^i$ and the corresponding gauge subgroup is $\caU_N(\algA) = \{ u \in M_N(\algA)\ / \ u^\invast u = u u^\invast = \bbbone_N\}$, the group of unitary elements of $M_N(\algA)$.

From the Serre-Swan theorem \cite{Serr55a, Swan62a}, it is well-known that a  vector bundle $\varE$ on a smooth manifold $\varM$ is completely characterized by its space of smooth sections $\Gamma(\varE)$ as a projective finitely generated right module over the commutative algebra $C^\infty(\varM)$. The natural generalization of a vector bundle in noncommutative geometry is then to take a projective finitely generated right $\algA$-module.

Let $\modM = p \algA^N$ be a projective finitely generated right $\algA$-module defined by a projection $p \in M_N(\algA)$ for some $N>0$. One can extend $p$ to a map $(\Omega^\grast)^N \rightarrow (\Omega^\grast)^N$ which acts on the left by matrix multiplication and $\modM \otimes_\algA \Omega^\grast = p (\Omega^\grast)^N$. If $\widehat{\nabla}^0$ is a noncommutative connection on the right module $\algA^N$, then $m \mapsto p \circ \widehat{\nabla}^0 m$ is a noncommutative connection on $\modM$, where $m \in \modM$ is considered as an element in $\algA^N$. In particular, $\widehat{\nabla}^0 m = \dd m$ is a natural connection on $\algA^N$ which defines the connection $\widehat{\nabla} m = p \circ \dd m$ on $\modM$. This connection depends only on the projection $p$. The curvature of this connection is the left multiplication on $\modM = p \algA^N \subset \algA^N$ by the matrix of $2$-forms $p \dd p \dd p$.

This construction shows that the space of noncommutative connections on a projective finitely generated right $\algA$-module is never empty. This is why we often assume to be in that situation to make sure we do not study an empty space. But as shown in the following situation, there are other situations where this space is clearly non empty.

For the last case we would like to describe, we suppose that the differential calculus has the following property: there exists a $1$-form $\xi \in \Omega^1$ such that $\dd a = [\xi, a]$ for any $a \in \algA$. Notice that we do not require this relation to hold in higher degrees in $\Omega^\grast$. Then for any right $\algA$-module $\modM$, the map $\modM \ni m \mapsto \widehat{\nabla}^{-\xi} m = - m \otimes \xi$ is a noncommutative connection on $\modM$. Indeed, one has $\widehat{\nabla}^{-\xi}(m a) = - m a \otimes \xi = - m \otimes a \xi = m \otimes [\xi, a] + m \otimes \xi a = (\widehat{\nabla}^{-\xi} m) a + m \otimes \dd a$ for any $m \in \modM$ and $a \in \algA$. The curvature of this canonical connection is $\widehat{R} m = m \otimes [\dd (-\xi) + (-\xi)(-\xi)]$ and using the defining property of $\xi$, one can show that $[\dd (-\xi) + (-\xi)(-\xi), a] = 0$ for any $a \in \algA$.

Let us restrict our analysis to the right $\algA$-module $\modM = \algA$. In that case, one has $\widehat{\nabla}^{-\xi} a = - a \xi$ for any $a \in \algA$. The connection $1$-form associated to this noncommutative connection is $\widehat{\nabla}^{-\xi} \bbbone = - \xi$. Let $g \in \algA$ be an invertible element, considered as a element of the gauge group. Then its action on this connection is given by $(-\xi)^g = - g^{-1} \xi g + g^{-1} \dd g = - g^{-1} \xi g + g^{-1} [\xi, g] = - \xi$ so that the connection $\widehat{\nabla}^{-\xi}$ is gauge invariant. Notice that the fact that the curvature $2$-form $\Omega = \dd (-\xi) + (-\xi)(-\xi)$ is gauge invariant is already known by the fact that $\Omega$ commutes with $\algA$.

This example is far from being academic: there are many examples of differential calculi satisfying this requirement. In that case, it is not necessary to demand that the module be projective and finitely generated to get a non empty space of noncommutative connections. Moreover, for the case $\modM = \algA$, this space of connections has an important point which is invariant by gauge transformations. This situation can only be encountered in noncommutative geometry, because in ordinary geometry any $1$-form commutes with the elements of the algebra.

%%%%%%%%%%%%%%%%%%%%%%%%%%%%%%%%%%%%%%%%%%%%%%%%%%%%%%%%%%%%%%%%%%%%%%%%%%
\section{Derivation-based noncommutative geometry}
\label{sec-Derivation-basednoncommutativegeometry}
%%%%%%%%%%%%%%%%%%%%%%%%%%%%%%%%%%%%%%%%%%%%%%%%%%%%%%%%%%%%%%%%%%%%%%%%%%

The derivation-based noncommutative geometry was initiated in \cite{DuVi88}. It has been exposed and studied for various algebras, for instance in \cite{DuViKernMado90a, DuViKernMado90b, Mass07, Mass11, Mass14, Mass15, DuViMich94, DuViMich96, DuViMich97, Mass32}. See \cite{DuVi01a, Mass30, Mass33} for reviews.

%%%%%%%%%%%%%%%%%%%%%%%%%%%%%%%%%%%%%%%%%%%%%%%%%%%%%%%%%%%%%%%
\subsection{Derivation-based differential calculus}
%%%%%%%%%%%%%%%%%%%%%%%%%%%%%%%%%%%%%%%%%%%%%%%%%%%%%%%%%%%%%%%

Let $\algA$ be an associative algebra with unit $\bbbone$. We denote by $\caZ(\algA) = \{ a \in \algA \ / \ ab = ba, \forall b \in \algA \}$ the center of $\algA$. The differential calculus we are interested in is constructed on the space of derivations of $\algA$, defined as
\begin{equation*}
\der(\algA) = \{ \kX : \algA \rightarrow \algA \ / \ \kX \text{ linear}, \kX\cdotaction(ab) = (\kX\cdotaction a) b + a (\kX\cdotaction b), \forall a,b\in \algA\}.
\end{equation*}
This vector space is a Lie algebra for the bracket $[\kX, \kY ]a = \kX  \kY a - \kY \kX a$ for all $\kX,\kY \in \der(\algA)$, and it is a $\caZ(\algA)$-module for the product $(f\kX )\cdotaction a = f(\kX\cdotaction a)$ for all $f \in \caZ(\algA)$ and $\kX \in \der(\algA)$.

The subspace $\Int(\algA) = \{ \adrep_a : b \mapsto [a,b]\ / \ a \in \algA\} \subset \der(\algA)$ is called the vector space of inner derivations. It is a Lie ideal and a $\caZ(\algA)$-submodule. We can define $\Out(\algA)=\der(\algA)/\Int(\algA)$ and we have the short exact sequence of Lie algebras and $\caZ(\algA)$-modules
\begin{equation*}
%\label{eq-secderivations}
\xymatrix@1@C=15pt{{\algzero} \ar[r] & {\Int(\algA)} \ar[r] & {\der(\algA)} \ar[r] & {\Out(\algA)} \ar[r] & {\algzero}}.
\end{equation*}
$\Out(\algA)$ is called the space of outer derivations of $\algA$. If $\algA$ is commutative, there are no inner derivations, so that the space of outer derivations is the space of all derivations.

In case $\algA$ has an involution, a derivation $\kX \in \der(\algA)$ is called real if $(\kX a)^\invast = \kX a^\invast$ for any $a\in \algA$. We denote by $\der_\gR(\algA)$ the space of real derivations.

We denote by $\underline{\Omega}^n_\der(\algA)$ the vector space of $\caZ(\algA)$-multilinear antisymmetric maps from $\der(\algA)^n$ to $\algA$, with $\underline{\Omega}^0_\der(\algA) = \algA$, and we define the total space
\begin{equation*}
\underline{\Omega}^\grast_\der(\algA) =\bigoplus_{n \geq 0} \underline{\Omega}^n_\der(\algA).
\end{equation*}
The space $\underline{\Omega}^\grast_\der(\algA)$ gets a structure of $\gN$-graded differential algebra for the product
\begin{multline*}
(\omega\eta)(\kX_1, \dots, \kX_{p+q}) = \\
 \frac{1}{p!q!} \sum_{\sigma\in \kS_{p+q}} (-1)^{\sign(\sigma)} \omega(\kX_{\sigma(1)}, \dots, \kX_{\sigma(p)}) \eta(\kX_{\sigma(p+1)}, \dots, \kX_{\sigma(p+q)})
\end{multline*}
for any $\kX_i \in \der(\algA)$. We define the differential $\hd$ by the so-called Koszul formula
\begin{multline*}
\hd\omega(\kX_1, \dots , \kX_{n+1}) = \sum_{i=1}^{n+1} (-1)^{i+1} \kX_i\cdotaction \omega( \kX_1, \dots \omi{i} \dots, \kX_{n+1}) \\[-5pt]
 + \sum_{1\leq i < j \leq n+1} (-1)^{i+j} \omega( [\kX_i, \kX_j], \dots \omi{i} \dots \omi{j} \dots , \kX_{n+1}). 
\end{multline*}
This formula is the one used to define the differential in complex spaces associated to Lie algebras.

Inside the differential calculus $(\underline{\Omega}^\grast_\der(\algA), \hd)$ lies a smaller one, defined as the sub differential graded algebra generated in degree $0$ by $\algA$. We denote it by $\Omega^\grast_\der(\algA) \subset \underline{\Omega}^\grast_\der(\algA)$. By definition, every element in $\Omega^n_\der(\algA)$ is a sum of terms of the form $a_0 \hd a_1 \cdots \hd a_n$ for $a_0, \dots, a_n \in \algA$. We will refer to $\underline{\Omega}^\grast_\der(\algA)$ as the maximal differential calculus and to $\Omega^\grast_\der(\algA)$ as the minimal one. 

Because the minimal differential calculus is generated by $\algA$, it is a quotient of the universal differential calculus, while the maximal differential calculus can contain elements which are not in this quotient.

The previous construction is motivated and inspired by the following situation. The algebra $\algA = C^\infty(\varM)$ of smooth functions on a smooth compact manifold $\varM$ is commutative, so that $\caZ(\algA) = C^\infty(\varM)$. It is well-known that $\der(\algA) = \Gamma(T\varM)$ is the Lie algebra of vector fields on $\varM$. Because the algebra is commutative, $\Int(\algA) = \algzero$, so that $\Out(\algA) = \Gamma(T\varM)$. The two graded differential algebras coincide with the graded differential algebra of de~Rham forms on $\varM$: $\Omega^\grast_\der(\algA) = \underline{\Omega}^\grast_\der(\algA) = \Omega^\grast(\varM)$.

%%%%%%%%%%%%%%%%%%%%%%%%%%%%%%%%%%%%%%%%%%%%%%%%%%%%%%%%%%%%%%%%%%%%%%%%%%
\subsection{Noncommutative connections}
%\label{}

As we will see, noncommutative connections constructed with the derivation-based differential calculus look very much like ordinary connections. 

Let $\modM$ be a right $\algA$-module. Explicitly, a noncommutative connection on $\modM$ for the differential calculus based on derivations is a linear map $\widehat{\nabla}_\kX : \modM \rightarrow \modM$, defined for any $\kX \in \der(\algA)$, such that for all $\kX,\kY \in \der(\algA)$, $a \in \algA$, $m \in \modM$, and $f \in \caZ(\algA)$ one has:
\begin{align*}
\widehat{\nabla}_\kX (m a) &=  (\widehat{\nabla}_\kX m) a + m(\kX\cdotaction a),
&
\widehat{\nabla}_{f\kX} m &= f \widehat{\nabla}_\kX m,
&
\widehat{\nabla}_{\kX + \kY} m &= \widehat{\nabla}_\kX m + \widehat{\nabla}_\kY m.
\end{align*}
The curvature of $\widehat{\nabla}$ identifies with the right $\algA$-module morphism $\hR(\kX, \kY) : \modM \rightarrow \modM$ defined for any $\kX, \kY \in \der(\algA)$ by $\hR(\kX, \kY) m = [ \widehat{\nabla}_\kX, \widehat{\nabla}_\kY ] m - \widehat{\nabla}_{[\kX, \kY]}m$.

As in ordinary geometry, it is possible to interpret the curvature as an obstruction on $\widehat{\nabla}$ to be a morphism of Lie algebras between $\der(\algA)$ and the space of (differential) operators on $\modM$. Notice that in ordinary geometry, we cannot make the distinction between the respective roles of the algebra and its center. Here it is essential to do a clear distinction between the two algebras, because $\der(\algA)$ is only a module over the center.

%%%%%%%%%%%%%%%%%%%%%%%%%%%%%%%%%%%%%%%%%%%%%%%%%%%%%%%%%%%%%%%%%%%%%%%%%%
\subsection{\texorpdfstring{The algebra $\algA = M_n(\gC)$}{The algebra A = Mn(C)}}
\label{subsec-thematrixalgebra}

Let us consider the case of the finite dimensional algebra $\algA = M_n(\gC) = M_n$ of $n\times n$ complex matrices. Its derivation-based differential calculus has been described in details in \cite{DuVi88, DuViKernMado90a, Mass11, Mass30}.

The center of $\algA$ is $\caZ(M_n) = \gC$. It is well-known that the matrix algebra has only inner derivations, and we have the identification $\der(M_n) = \Int(M_n) \simeq \ksl_n =\ksl_n(\gC)$ where $\ksl_n(\gC)$ is the $n^2-1$-dimensional Lie algebra of traceless complex $n \times n$ matrices. The explicit isomorphism associates to any $\gamma \in \ksl_n$ the derivation $\adrep_\gamma : a \mapsto [\gamma, a]$. Because $\der(\algA) = \Int(\algA)$, one has $\Out(M_n) = \algzero$: this is the opposite situation to the one encountered for commutative algebras.

For the involution given by adjointness, the space of real derivations is $\der_\gR(M_n) = \ksu(n)$, the Lie algebra of traceless Hermitian matrices. The identification is given explicitly by $\gamma \mapsto \adrep_{i \gamma}$ for any $\gamma \in \ksu(n)$. An explicit decription of the associated derivation-based differential calculus shows that 
\begin{equation*}
\underline{\Omega}^\grast_\der(M_n) = \Omega^\grast_\der(M_n) \simeq M_n \otimes \exter^\grast \ksl_n^\dualast,
\end{equation*}
with a differential, denoted by $\dd'$ in the following, which identifies with the differential of the Chevalley-Eilenberg complex of the Lie algebra $\ksl_n$ represented on $M_n$ by the adjoint representation (commutator) \cite{CartEile56, Weib97, Mass31}. In particular, the maximal and minimal differential calculi coincide, and we will use the notation $\Omega^\grast_\der(M_n)$ to designate it.

The canonical noncommutative $1$-form $i\theta \in \Omega^1_\der(M_n)$ defined, for any $\gamma \in M_n(\gC)$, by
\begin{equation*}
i\theta(\adrep_{\gamma}) = \textstyle\gamma - \frac{1}{n} \tr (\gamma)\bbbone,
\end{equation*}
makes the explicit isomorphism $\Int(M_n) \xrightarrow{\simeq} \ksl_n$. Moreover, it satisfies $\dd' a = [i\theta, a] \in \Omega^1_\der(M_n)$ for any $a \in M_n$. This relation is no more true in higher degrees. The differential of $i\theta$ is non zero, and one has $\dd' (i\theta) - (i\theta)^2 = 0$, which makes $i\theta$ looks very much like the Maurer-Cartan form in the geometry of Lie groups (here $SL_n(\gC)$).

In order to perform explicit computations, it is convenient to introduce a particular basis on this algebra. We denote by $\{E_k\}_{k=1, \dots, n^2-1}$ a basis of $\ksl_n$ of traceless Hermitian matrices. These elements define a basis for the Lie algebra $\der(M_n) \simeq \ksl_n$ through the $n^2-1$ (real) derivations $\partial_k = \adrep_{i E_k}$. Adjoining the unit $\bbbone$ to the $E_k$'s, one gets a basis for $M_n$. Obviously, the unit does not give rise to a derivation. We denote by $C^m_{k \ell}$ the real structure constants of $\ksl_n$ in this basis: $[E_k, E_\ell] = -i C^m_{k \ell} E_m$, so that $[\partial_k, \partial_\ell] = C^m_{k \ell} \partial_m$.

We introduce the dual basis $\{\theta^\ell\}$ in $\ksl_n^\dualast$ by $\theta^\ell(\partial_k) = \delta^\ell_k$. This basis generates a basis for the exterior algebra $\exter^\grast \ksl_n^\dualast$, where by definition one has $\theta^\ell \theta^k = - \theta^k \theta^\ell$.

Any noncommutative $p$-form ii explicitly decomposed as a sum of terms of the form $a \otimes \theta^{k_1} \cdots \theta^{k_p}$ for $k_1 < \cdots < k_p$ and for $a = a^k E_k + a^0 \bbbone \in M_n$. Using the derivation rule of $\dd'$, an explicit description of the differential $\dd'$ is given once we know it on the generators in degrees $0$ and $1$, for which one has $\dd' \bbbone = 0$, $\dd' E_k = -C^m_{k\ell} E_m \otimes \theta^\ell$, and $\dd' \theta^k = -\tfrac{1}{2} C^k_{\ell m} \theta^\ell \theta^m$.

The noncommutative $1$-form $i\theta$ can be written as $i\theta = i E_k \otimes \theta^k \in M_n \otimes \exter^1 \ksl_n^\dualast$. This relation is obviously independent of the chosen basis.

For any $\gamma, \eta \in \ksl_n \simeq \der(M_n)$, let us define $g(\gamma, \eta) = \frac{1}{n} \tr(\gamma\eta)$, which induces a natural non degenerated scalar product on $\der(M_n)$. Consider now the symmetric matrix $g_{k \ell} = \frac{1}{n} \tr(E_k E_\ell)$. These coefficients plays the role of a metric on the noncommutative space $M_n$, to which one can associate a Hodge star operation as follows. This is a map $\hodgeast : \Omega^p_\der(M_n) \rightarrow \Omega^{n^2-1-p}_\der(M_n)$ defined by 
\begin{equation*}
\hodgeast (a \otimes \theta^{k_1} \cdots \theta^{k_p}) = \frac{1}{(n^2-1-p)!} \sqrt{|g|} g^{k_1 \ell_1} \cdots g^{k_p \ell_p} \epsilon_{\ell_1 \ldots \ell_{n^2-1}} a \otimes \theta^{\ell_{p+1}} \cdots \theta^{\ell_{n^2-1}}
\end{equation*}
where $|g|$ is the determinant of the matrix $(g_{k \ell})$ and $\epsilon_{\ell_1 \ldots \ell_{n^2-1}}$ is the completely antisymmetric tensor.

There is a natural integration on the space of forms of maximal degree. Every differential $(n^2-1)$-form $\omega \in \Omega^{n^2-1}_\der(M_n)$ can be written uniquely as $\omega = a \sqrt{|g|} \theta^1 \cdots \theta^{n^2-1}$, where $a \in M_n$. The quantity $\sqrt{|g|} \theta^1 \cdots \theta^{n^2-1}$ depends only on the choice of an orientation on the basis $\{ \theta^k \}$ which we fix once and for all. Then the coefficient $a$ does not depend of the basis. We define the noncommutative integration of a noncommutative form $\omega$ as a map
\begin{equation*}
\int_{\text{n.c.}} : \Omega^\grast_\der(M_n) \rightarrow \gC,
\end{equation*}
given by $\int_{\text{n.c.}} \omega = \frac{1}{n} \tr(a)$ when $\omega \in \Omega^{n^2-1}_\der(M_n)$ is written as above, and $0$ otherwise. This integration satisfies the closure relation
\begin{equation*}
\int_{\text{n.c.}} \dd' \omega = 0.
\end{equation*}

As a first example of noncommutative gauge theory in this context, let us consider the simple case of the right $\algA$-module $\modM = \algA$. As we will see, this situation is not trivial.

The $1$-form $i\theta$ satisfies the requirement of the last example presented in \ref{sec-examples-ncconnections}. We then associate to it the canonical noncommutative connection given by $\widehat{\nabla}^{-i\theta}_\kX a = - a i\theta(\kX) = \kX\cdotaction a - i\theta(\kX) a = -a \gamma$ for any $a \in \algA$ and any $\kX = \adrep_\gamma \in \der(M_n)$ (with $\tr \gamma = 0$). 

This noncommutative connection is gauge invariant and its curvature is zero because it coincides with the multiplication by the $2$-form $\dd (-i\theta) + (-i\theta)(-i\theta) = 0$. It is a particular and preferred element in the affine space of noncommutative connection along which one can decompose any noncommutative connection as
\begin{equation*}
\widehat{\nabla}_\kX a = \widehat{\nabla}^{-i\theta}_\kX a + A(\kX) a = (A - i\theta)(\kX) a,
\end{equation*}
for a noncommutative $1$-form $A = A_k \otimes \theta^k \in \Omega^1_\der(M_n)$. Such a connection is compatible with the natural Hermitian structure $\langle a, b \rangle = a^\invast b$ on the module $\algA$ if and only if $A(\kX)^\invast  = -A(\kX)$ for any real derivation $\kX$. The $\partial_k$'s being real, this is equivalent to $A_k$ to be a anti-Hermitian matrix, which we assume in the following.

Under a gauge transformation $g \in U(n)$ compatible with the Hermitian structure, one has $A_k \mapsto g^{-1} A_k g$: the inhomogeneous term has been absorbed by $-i\theta$.

A straightforward computation shows that the curvature of $\widehat{\nabla}$ is the multiplication on the left by the $2$-form
\begin{equation*}
F = \tfrac{1}{2}( [A_k, A_\ell] - C^m_{k \ell} A_m) \otimes \theta^k \theta^\ell.
\end{equation*}
The matrices $F_{k \ell} = [A_k, A_\ell] - C^m_{k \ell} A_m$ are anti-Hermitian.

The natural action functional for this connection is 
\begin{equation*}
S[A] = \frac{1}{2} \int_{\text{n.c.}} F^\invast \, \hodgeast F = - \tfrac{1}{8n}  \tr \left( F_{k \ell}F^{k \ell}\right)
\end{equation*}
where $F^\invast$ is the involution applied to the anti-Hermitian $2$-form $F$. One has $S[A] \geq 0$ and the minimum is obtained in two situations: $\widehat{\nabla}$ is a pure gauge connection or $\widehat{\nabla} = \widehat{\nabla}^{-i\theta}$ is the canonical gauge invariant connection.

This gauge theory can be generalized using a right $\algA$-module of the form $\modM = M_{r,n}$, the vector space of $r\times n$ complex matrices with the obvious right $M_n$-module structure and the natural Hermitian structure given by $\langle m_1, m_2 \rangle = m_1^\ast m_2 \in M_n$.

The noncommutative connection $\widehat{\nabla}^{-i\theta}_\kX m = - m i\theta(\kX)$ is well defined, it is compatible with the Hermitian structure and its curvature is zero. We can use it to decompose any noncommutative connection as $\widehat{\nabla}_\kX m = \widehat{\nabla}^{-i\theta}_\kX m + A(\kX) m$ for $A = A_k \otimes \theta^k$ with $A_k \in M_r$. The curvature of $\widehat{\nabla}$ is the multiplication on the left by the $M_r$-valued noncommutative $2$-form $F = \frac{1}{2}( [A_k, A_\ell] - C^m_{k \ell} A_m) \otimes \theta^k \theta^\ell$. This curvature vanishes if and only if $A : \ksl_n \rightarrow M_r$ is a representation of the Lie algebra $\ksl_n$. Two flat connections are in the same gauge orbit if and only if the corresponding Lie algebra representations are equivalent. For more details, we refer to \cite{DuViKernMado90a}.

%%%%%%%%%%%%%%%%%%%%%%%%%%%%%%%%%%%%%%%%%%%%%%%%%%%%%%%%%%%%%%%%%%%%%%%%%%
\subsection{\texorpdfstring{The algebra $\algA = C^\infty(\varM) \otimes M_n(\gC)$}{The algebra A = Cinfty(M) x Mn(C)}}
\label{subsec-algebrafunctionsmatrices}

This noncommutative geometry is interesting because it mixes an ordinary geometry with the purely algebraic structure studied in the previous example. The algebra we consider is the space of smooth applications from a $m$-dimensional compact smooth manifold $\varM$ into the matrix algebra $M_n$. The algebra identifies with the tensor product $\algA = C^\infty(\varM) \otimes M_n(\gC)$. The derivation-based differential calculus of this algebra was first considered in \cite{DuViKernMado90b}, to which we refer for further details.

The center of this algebra is the purely geometric part $\caZ(\algA) = C^\infty(\varM)$, where we identify a function $f \in C^\infty(\varM)$ with the application $f \bbbone_n$, where $\bbbone_n$ is the identity matrix in $M_n$.

The space of derivations of $\algA$ can be decomposed into two parts as $\der(\algA) = [\der(C^\infty(\varM))\otimes \bbbone_n ] \oplus [ C^\infty(\varM) \otimes \der(M_n) ] = \Gamma(T\varM) \oplus [C^\infty(\varM) \otimes \ksl_n]$. Using the notation $\kX = X \oplus \gamma \in \der(\algA)$, with $X \in \Gamma(T\varM)$ and $\gamma : \varM \rightarrow\ksl_n$, the Lie structure is given by $[\kX, \kY] = [X,Y] \oplus (X \cdotaction \eta - Y \cdotaction \gamma + [\gamma, \eta])$ for $\kY = Y \oplus \eta \in \der(\algA)$ where $X\cdotaction \eta$ is the action of $X$ as a vector field on the map $\eta$.

We denote by $\algA_0 = C^\infty(\varM) \otimes \ksl_n$ the Lie algebra of traceless elements in $\algA$ for the commutator of matrices. Then $\algA_0 = \Int(\algA)$. We can identify $\Out(\algA) = \Gamma(\varM)$. 

The maximal and minimal differential calculi coincide and, using the decomposition of $\der(\algA)$, they can be identified with the tensor product  of the de~Rham differential calculus on $\varM$ with the differential calculus on the matrix algebra:
\begin{equation*}
\underline{\Omega}^\grast_\der(\algA) = \Omega^\grast_\der(\algA) = \Omega^\grast(\varM) \otimes \Omega^\grast_\der(M_n).
\end{equation*}
The differential is the sum $\hd = \dd + \dd'$, where $\dd$ is the de~Rham differential and $\dd'$ is the differential introduced in the previous example.

The noncommutative $1$-form $i\theta$ defined by $i\theta(X \oplus \gamma) = \gamma$ gives explicitly the split of the short exact sequence of Lie algebras and $C^\infty(\varM)$-modules given by the quotient of $\der(\algA)$ by $\Int(\algA)$:
\begin{equation}
\label{eq-splittingsecderivationstrivialcase}
\xymatrix@1@C=25pt{{\algzero} \ar[r] & {\algA_0} \ar[r] & {\der(\algA)} \ar[r] \ar@/_0.7pc/[l]_-{i\theta}& {\Gamma(\varM)} \ar[r] & {\algzero}}.
\end{equation}

The noncommutative integration defined on $\Omega^\grast_\der(M_n)$ in the previous example extends to a well-defined map of differential complexes
\begin{align*}
\int_{\text{n.c.}} : \Omega^\grast_\der(\algA) & \rightarrow \Omega^{\grast - (n^2-1)}(\varM),
&
\int_{\text{n.c.}} \hd \omega &= \dd \int_{\text{n.c.}} \omega.
\end{align*}

Using a metric $h$ on $\varM$ and the metric $g_{k \ell} = \frac{1}{n} \tr(E_k E_\ell)$ on the matrix part, one can define a metric on $\der(\algA)$ by $\widehat{g}(X+ \adrep_\gamma, Y + \adrep_\eta) = h(X,Y) + \textstyle \frac{1}{\mu^2}g(\gamma \eta)$ where $\mu$ is a positive constant which measures the relative ``weight'' of the two ``spaces''. In physical natural units, it has the dimension of a mass.

This metric defines a Hodge star operator $\hodgeast : \Omega^p_\der(\algA) \rightarrow \Omega^{m+n^2-1-p}_\der(\algA)$ which can be obtained either by a direct construction performed in a basis of $\der(\algA)$ using the metric $\widehat{g}$, either by the composition of the two Hodge star operations associated to $h$ and $g$ respectively on the first and second factor of $\Omega^\grast(\varM) \otimes \Omega^\grast_\der(M_n)$. A scalar product can then be defined on $\Omega^\grast_\der(\algA)$ by
\begin{equation*}
( \omega, \eta) =\int_\varM \int_{\text{n.c.}} \omega^\invast\, \hodgeast \eta,
\end{equation*}
where $\omega \mapsto \omega^\invast$ is the natural involution induced on noncommutative forms by the involution on $M_n$.

Let us describe the gauge theory associated to the right $\algA$-module $\modM = \algA$. For the algebra $M_n$, the noncommutative $1$-form $-i\theta$ defines a canonical noncommutative connection by the relation $\widehat{\nabla}^{-i\theta}_\kX a = \kX\cdotaction a - i\theta(\kX) a$ for any $a \in \algA$. In the present situation, we can use again this connection as a particular (and canonical) one.

For any $a \in \algA$ and $\kX = X \oplus \gamma \in \der(\algA)$, this connection takes the explicit form $\widehat{\nabla}^{-i\theta}_\kX a = X \cdotaction a - a \gamma$. Its curvature is zero. But this connection is no more gauge invariant, and the connection $1$-form associated to the transformed connection $(\widehat{\nabla}^{-i\theta})^g$ by $g \in C^\infty(\varM) \otimes GL_n(\gC)$ is given by $\kX \mapsto -i\theta(\kX) + g^{-1} (X\cdotaction g) = - \gamma + g^{-1} (X\cdotaction g)$.

Let $\widehat{\nabla}$ be a noncommutative connection on $\algA$, written as $\widehat{\nabla}_\kX a = \kX \cdotaction a + \omega(\kX) a$ where $\omega = \widehat{\nabla} \bbbone$ is its associated connection $1$-form. We can also decompose $\widehat{\nabla}$ as $\widehat{\nabla}_\kX a = \widehat{\nabla}^{-i\theta}_\kX a + A(\kX) a$ where $A \in \Omega^1_\der(\algA)$ is related to $\omega$ by $A = \omega + i\theta$. Let us decompose $A$ as $A(X \oplus \gamma) = \ka(X) + \kb(\gamma)$ for $\ka = \ka_\mu \dd x^\mu \in M_n \otimes \Omega^1(\varM)$ and $\kb = \kb_k \theta^k \in C^\infty(\varM) \otimes M_n \otimes \exter^1 \ksl_n^\dualast$.

The compatibility of $\widehat{\nabla}$ with the natural Hermitian structure on the right module $\algA$ implies that $A$ takes its values in anti-Hermitian matrices, which we suppose in the following.

Under a gauge transformation $g : \varM \rightarrow U(n)$ compatible with the Hermitian structure, one has
\begin{align*}
\ka_\mu &\mapsto g^{-1} \ka_\mu g + g^{-1} \partial_\mu g,
&
\kb_k &\mapsto g^{-1} \kb g.
\end{align*}
The curvature of $\widehat{\nabla}$ is then the noncommutative $2$-form
\begin{multline*}
F = \textstyle\frac{1}{2}(\partial_\mu \ka_\nu - \partial_\nu \ka_\mu + [ \ka_\mu, \ka_\nu])\dd x^\mu \dd x^\nu 
\\
 + (\partial_\mu \kb_k + [ \ka_\mu, \kb_k]) \dd x^\mu \theta^k + \frac{1}{2}([\kb_k, \kb_\ell] - C^m_{k \ell} \kb_m) \theta^k \theta^\ell.
\end{multline*}

Using the metric on $\der(\algA)$ (where $h$ is taken to be euclidean), the associated Hodge star operation on forms, and the scalar product on forms, one can define the following action functional associated to the connection $\widehat{\nabla}$:
\begin{multline*}
S[A] = -\frac{1}{4n} \int \dd x \tr\bigg\{ \sum_{\mu,\nu}(\partial_\mu \ka_\nu - \partial_\nu \ka_\mu + [ \ka_\mu, \ka_\nu])^2 \\[-10pt]
-\frac{\mu^2}{2n} \sum_{\mu,k} (\partial_\mu \kb_k + [ \ka_\mu, \kb_k])^2 
-\frac{\mu^4}{4n} \sum_{k, \ell} ([\kb_k, \kb_\ell] - C^m_{k \ell} \kb_m)^2 \bigg\}.
\end{multline*}
The integrand is zero on two gauge orbits. The first one corresponds to $\ka = g^{-1} \dd g$ and $b_k=0$. It is the gauge orbit of $\widehat{\nabla} = \widehat{\nabla}^{-i\theta}$. The second one corresponds to $\ka_\mu=g^{-1} \dd g$ and $b_k= i g^{-1} E_k g$. It is the gauge orbit of $\widehat{\nabla}_\kX a = \kX\cdotaction a$ ($\omega = 0$).

The configurations with $b_k = i E_k$ describe connections where the $a_\mu$ have a mass term coming from the second term of this Lagrangian. This is a Higgs-like mechanism where the scalar fields are the $b_k$ fields, coupled to the $U(n)$-Yang-Mills fields $a_\mu$ through a covariant derivative in the adjoint representation. These fields are not introduced by hand in the Lagrangian: they are part of the noncommutative connection along the purely algebraic directions. This Yang-Mills-Higgs model is very constrained and does not allow for any arbitrariness.

As for the matrix algebra, one can consider a higher rank right $\algA$-module of the form $\modM = C^\infty(\varM) \otimes M_{r,n}$. Then, performing a similar analysis, one gets the following main features of the associated gauge field theory: there are non trivial flat connections and they are classified by inequivalent representations of $\ksl_n$ in $M_r$ \cite{DuViKernMado90b}.

The algebra $\algA = C^\infty(\varM) \otimes M_n(\gC)$ admits a natural generalization in terms of fiber bundle theory. Let $\varP$ be a $SU(n)$-principal fiber bundle and let $\varE$ be the associated vector bundle for the fundamental representation of $SU(n)$ on $\gC^n$. Denote by $\algA$ the associative algebra of smooth sections of the vector bundle $\varE \otimes \varE^\dualast$, whose fiber is $M_n(\gC)$. This is the algebra of endormorphisms of $\varE$. The particular trivial situation $\varP = \varM \times SU(n)$ gives rise to $\algA = C^\infty(\varM) \otimes M_n(\gC)$ while a more general situation can take into account the non triviality of $\varP$. 

The derivation-based noncommutative geometry of this algebra has been studied in \cite{Mass14, Mass15, Mass25}, see \cite{Mass30} for a review. The main difference between the general case and the trivial situation lies in the short exact sequence of Lie algebras and $C^\infty(\varM)$-modules \eqref{eq-splittingsecderivationstrivialcase} which splits in the trivial case but does not split for non trivial fiber bundles. A splitting $\nabla : \Gamma(T\varM) \rightarrow \algA$ of this short exact sequence as $C^\infty(\varM)$-modules is given by an ordinary connection on $\varP$ ($\nabla$ is the associated covariant derivative on $\varE \otimes \varE^\dualast$). Concerning gauge field theories, many of the features shown before for the trivial situation remain valid, modulo that they have to be adapted to a non trivial global topology.

The main advantage of this noncommutative geometry is that it permits to embed the space of ordinary connections on $\varP$ into the space of noncommutative connections on $\algA$. In this embedding, the corresponding notions of curvatures and of gauge transformations are in correspondance. We refer to \cite{Mass30} for more details.

%%%%%%%%%%%%%%%%%%%%%%%%%%%%%%%%%%%%%%%%%%%%%%%%%%%%%%%%%%%%%%%%%%%%%%%%%%
\subsection{The Moyal algebra}
%\label{}

Field theories on the Moyal algebra have been extensively studied since the discovery of a modified $\phi^4$ theory which is renormalizable to all orders \cite{GrosWulk05b, GrosWulk05a}, see \cite{RIVASSEAU2007HAL-001656861, WALLET2007HAL-001709651} for reviews. The Moyal algebra gives rise to gauge field theories as well, whose content depends explicitly on the choice of the differential calculus \cite{Mass32}.

Let us recall the definition of the Moyal algebra, restricted here to the $2$-dimensional case. This is a deformation of the algebra of smooth functions on the plane $\gR^2$. Different Moyal algebras can be defined \cite{GraciaBondia1987kw,Varilly1988jk,Gayral2003dm}. We will choose here the one commonly used in noncommutative field theories, often called the ``Moyal multiplier algebras''.

Let $\caS(\gR^2)$ be the space of complex-valued Schwartz functions on the plane $\gR^2$, and let $\caS'(\gR^2)$ be the space of associated tempered distributions. Let $\Theta = \theta \spmatrix{0 & -1 \\ 1 & 0}$ be an antisymmetric matrix, with $\theta \in \gR$, $\theta \neq 0$, the deformation parameter. The Moyal-Groenenwald product $\caS(\gR^2) \times \caS(\gR^2) \rightarrow \caS(\gR^2)$ is defined by the integral formula
\begin{equation*}
(f \moyast g)(x)=\frac{1}{(\pi\theta)^2}\int d^2 y d^2 z\ f(x+y) g(x+z) e^{-i2 y \Theta^{-1} z}. 
\end{equation*}
This product is extended to give a left and a right module structures on $\caS'(\gR^2)$ by the relations
\begin{align*}
\caS(\gR^2) \times \caS'(\gR^2) &\rightarrow \caS'(\gR^2)
&
\caS'(\gR^2) \times \caS(\gR^2) &\rightarrow \caS'(\gR^2)
\\
\langle f \moyast T, g \rangle &= \langle T, g \moyast f \rangle
&
\langle T \moyast f, g \rangle &= \langle T, f \moyast g \rangle
\end{align*}
where $\langle T , f \rangle$ is the coupling between $\caS'(\gR^2)$ and $\caS(\gR^2)$. The smoothening property of the Moyal product ensures that $f \moyast T$ and $T \moyast f$ are smooth functions. We then define the left and right multiplier spaces by $\caL = \{ T \in \caS'(\gR^2) \ / \ f \moyast T \in \caS(\gR^2), \forall f \in \caS(\gR^2) \}$ and $\caR = \{ T \in \caS'(\gR^2) \ / \ T \moyast f \in \caS(\gR^2), \forall f \in \caS(\gR^2) \}$.

The Moyal algebra is defined by $\algA_\Theta = \caL \cap \caR$. This algebra contains $\caS(\gR^2)$ as an ideal, and it contains also the polynomials functions on $\gR^2$. For the particular coordinate polynomials $x^\mu$ one has $[x^\mu,x^\nu]_\moyast = i\Theta^{\mu\nu}$, which is often taken as the heuristic starting point to define the Moyal algebra.

The center of $\algA_\Theta$ is $\caZ(\algA_\Theta) = \gC$ and all the derivations are inner: $\der(\algA_\Theta) = \Int(\algA_\Theta)$. For instance, the usual partial derivative on functions is a derivation on $\algA_\Theta$ and it can be written as $\partial_\mu a = [ -i\Theta^{-1}_{\mu\nu} x^\nu, a]_\moyast$ for any $a \in \algA_\Theta$.

The Lie algebra of derivations of $\algA_\Theta$ is an infinite dimensional vector space. Using this full space of derivations to construct gauge field theories would lead to gauge potentials with an infinite number of field components. This is why we will introduce a differential calculus based on a finite dimensional Lie sub algebra of $\der(\algA_\Theta)$. But then the choice is not canonical, and it leads to different gauge field theories.

An implicit natural choice has been made in the literature \cite{DEGOURSAC2007HAL-001359171, Grosse2006hh, Grosse2007dm, deGoursac2008rb, WALLET2007HAL-001709651}: it consists to consider the $2$-dimensional Lie algebra with basis $\{\partial_\mu\}_{\mu=1,2}$. Then, they are two gauge potentials $\{A_\mu\}_{\mu=1,2}$, defined by $\widehat{\nabla}_{\partial_\mu} \bbbone = A_\mu$, where, for simplicity, we consider only the case of the right
$\algA_\Theta$-module $\algA_\Theta$ itself.

A second choice has been studied in \cite{Mass32} (see \cite{Mass33} for a review and a more complete discussion). It consists to consider the $5$-dimensional Lie subalgebra $\kisp(2,\gR)^\gC$ (complexified Lie algebra of the inhomogeneous symplectic Lie algebra on $\gR^2$), which, acting on the Moyal algebra, is the space of inner derivations coming from polynomial functions of degree less than or equal to $2$. 

$\kisp(2,\gR)^\gC$ is the maximal Lie subalgebra of derivations of $\algA_\Theta$ which are also derivations of the ordinary (commutative) algebra generated by $\caS(\gR^2)$ and polynomial functions. As a consequence, the directions defined by these derivations have clear geometrical interpretations: the two derivations $\partial_\mu$, $\mu=1,2$, are associated to ordinary translations, and they will be called ``spatial'' directions, while the three others are associated to symplectic rotations (for the symplectic $2$-form $\Theta$). See \cite{Mass32} for the explicit construction.

We will denote by $(\Omega_\kisp^\grast(\algA_\Theta), d)$ the derivation-based differential calculus constructed on this Lie subalgebra. We summarize here the results obtained in \cite{Mass32} about gauge field theories constructed with these structures.

There is a canonical noncommutative $1$-form $\eta \in \Omega_\kisp^1(\algA_\Theta)$ such that $da = [\eta, a]$. It is defined by $\eta(\adrep_P) = P_0$ where, for any polynomial function $P$ of degree less than or equal to $2$, $P_0$ is the polynomial function $P$ from which we remove the constant part, which is in the center of $\algA_\Theta$. Thus $\eta(\partial_\mu) = -i\Theta^{-1}_{\mu\nu} x^\nu$ for the spatial directions.

As in \ref{sec-examples-ncconnections}, this noncommutative $1$-form defines the canonical noncommutative connection $\widehat{\nabla}_\kX a = - a \eta(\kX)$ for any $a \in \algA_\Theta$ and any $\kX \in \kisp(2,\gR)^\gC$. It curvature is $\hR(\kX, \kY) a = a \left( \eta([\kX, \kY]) - [\eta(\kX), \eta(\kY)] \right)$. We saw in \ref{sec-examples-ncconnections} that the $2$-form $(\kX, \kY) \mapsto \Omega(\kX,\kY) = \eta([\kX, \kY]) - [\eta(\kX), \eta(\kY)]$ is necessary in the center of $\algA_\Theta$, so that it takes its values in $\gC$. In an explicit basis of $\kisp(2,\gR)^\gC$, $\Omega(\kX,\kY)$ is non zero only in the spatial directions, where it takes values proportional to $\Theta^{-1}_{\mu\nu}$.  Contrary to the canonical noncommutative $1$-form $i\theta \in \Omega^1_\der(M_n)$ introduced in \ref{subsec-thematrixalgebra}, $\eta$ cannot be defined to be a morphism of Lie algebras, and the curvature $\Omega$ measures this failure.

In \ref{subsec-algebrafunctionsmatrices} we saw that the purely noncommutative directions (the directions along $\ksl_n$) of the gauge potentials can be interpreted as Higgs fields. This relied heavily on the fact that the canonical connection defined by $i\theta$ is of zero curvature. The present situation is quite similar. The curvature is zero in the directions of the symplectic rotations, and it is shown in \cite{Mass32} that the gauge potentials in these directions can indeed been interpreted as Higgs fields. Comparing the two situations, one is tempted to interpret the symplectic rotations as ``inner'' symmetries and the spatial directions as ``outer'' ones. But keep in mind that all derivations are inner in the algebraic sense.

%%%%%%%%%%%%%%%%%%%%%%%%%%%%%%%%%%%%%%%%%%%%%%%%%%%%%%%%%%%%%%%%%%%%%%%%%%
\section{Spectral triples}
%\label{}
%%%%%%%%%%%%%%%%%%%%%%%%%%%%%%%%%%%%%%%%%%%%%%%%%%%%%%%%%%%%%%%%%%%%%%%%%%

Gauge theories based on spectral triples use a different approach to the one exposed in the previous section, in particular concerning the origin of the gauge transformations. Nevertheless, differential calculus and module structures play an essential role. For some reviews of this approach, see for instance \cite{Conn94, MR2179016, MR2371808,ScheHaraWern02a}.

%%%%%%%%%%%%%%%%%%%%%%%%%%%%%%%%%%%%%%%%%%%%%%%%%%%%%%%%%%%%%%%%%%%%%%%%%%
\subsection{The axioms}
%\label{}

A spectral triple is nowadays more than a triple, but it is based on three essential components which correspond indirectly to the ones introduced to define a noncommutative connection. 

A triple spectral $(\algA, \ehH, \caD)$ is given by a unital $C^\ast$-algebra $\algA$, a faithful involutive representation $\pi : \algA \rightarrow \caB(\ehH)$ on a Hibert space $\ehH$ and an unbounded self-adjoint operator $\caD$ on $\ehH$, called a Dirac operator, such that:
\begin{itemize}
\item the set $\caA = \{ a \in \algA \ / \ [\caD, \pi(a)] \text{ is bounded} \}$ is norm densed in $\algA$;
\item $(1+\caD^2)^{-1}$ has compact resolvent.
\end{itemize}

The representation makes $\ehH$ into a left $\algA$-module. The Dirac operator $\caD$ is used to defined a differential structure. The sub algebra $\caA$ identifies with the ``smooth functions'' on the (noncommutative) space and the differential of $a \in \caA$ is more or less $\dd a = [\caD, a]$ (more on this latter). This is just an heuristic formula: the commutator with $\caD$ cannot be used to define a true differential.

The decreasing rate of the eigenvalues of $|\caD|^{-1}$ defines an integer $n$ associated to the spectral triple, which is called its dimension.

The Dirac operator gives also a geometric structure to the spectral triple, in the sense of a way to measure ``lengths'' (between states). This is out of the scope of this review to further develop this side of the theory (see \cite{Conn94} for further details).

A spectral triple is said to be even when $n$ is even and when there exists an operator $\gamma : \caH \rightarrow \caH$ such that $\gamma^\invast = \gamma$, $\caD \gamma + \gamma \caD =0$, $\gamma \pi(a) - \pi(a) \gamma =0$, and $\gamma^2 = 1$, for any $a \in \algA$. The operator $\gamma$ is called a chirality.

A spectral triple is said to be real when there exists an antiunitary operator $J : \caH \rightarrow \caH$ such that $[J \pi(a) J^{-1}, \pi(b)] =0$, $J^2 = \epsilon$, $J \caD = \epsilon' \caD J$ and $J \gamma = \epsilon'' \gamma J$ for any $a,b \in \algA$. The coefficients $\epsilon, \epsilon'$, and $\epsilon''$ take their values in the following table, which depends on the dimension $n$ of the spectral triple:
\begin{center}
\begin{tabular}{*{9}{>{$}r<{$}}}
\toprule
n \mod 8   & 0 &  1 &  2 &  3 &  4 &  5 &  6 & 7 \\
\midrule
\epsilon   & 1 &  1 & -1 & -1 & -1 & -1 &  1 & 1 \\
\epsilon'  & 1 & -1 &  1 &  1 &  1 & -1 &  1 & 1 \\
\epsilon'' & 1 &    & -1 &    &  1 &    & -1 &   \\
\bottomrule
\end{tabular}
\end{center}

The map $a \mapsto J \pi(a)^\invast J^{-1} \in \caB(\ehH)$ is an involutive representation of $\algA$, and by definition it commutes with the representation $\pi$. This induces a structure of bimodule on $\ehH$ which plays an essential role in the following. We denote it by $(a,b) \mapsto \pi(a) J \pi(b)^\invast J^{-1}\Psi \simeq \pi(a) \Psi \pi(b)$ for any $\Psi \in \ehH$. Notice that the presence of $J$ in this formula implies the use of $\pi(b)^\invast$ instead of $\pi(b)$. The operator $\caD$ is required to be a first order differential operator for this bimodule structure \cite{Mass08}, which means concretely that $\left[ [\caD, \pi(a)], J \pi(b) J^{-1}  \right] = 0$ for any $a,b \in \algA$.

This restricted list of axioms for a spectral triple is sufficient to understand the principles of the gauge theories constructed in the following.

As an example, let us consider the commutative prototype of a spectral triple. Let $\varM$ be a smooth compact Riemannian spin manifold of dimension $m$. The algebra is $\algA = C(\varM)$, the commutative algebra of continuous functions on $\varM$. The Hilbert space is $\ehH = L^2(\slashed{S})$, where $\slashed{S}$ is a spin bundle given by the spin structure on $\varM$. The Dirac operator $\caD = \slashed{\partial}$ is a Dirac operator on $\slashed{S}$ associated to the Levi-Civita connection. For this spectral triple, the dimension is $m$ and the sub algebra $\caA$ is $C^\infty(\varM)$. If $m$ is even, then the chirality is given by $\gamma_\varM = - \gamma^1 \gamma^2 \cdots \gamma^m$ with $\gamma^\mu$ the Dirac gamma matrices satisfying $\{\gamma^\mu, \gamma^\nu\} = 2 g^{\mu \nu}$. Finally, the charge conjugation defines a real structure $J_\varM$ on this spectral triple.

A spectral triple can be identified with an unbounded Fredholm module, and it always defines a class in $K$-homology. For spaces with involution, the correct version of $K$-homology is $KR$-homology, and its dual is $KR$-theory, which is the $K$-theory of real vector bundles in the sense of ``spaces with involution'' \cite{Atiy66a}. A real spectral triple defines a class in $KR$-homology. The table of the coefficients $\epsilon$, $\epsilon'$ and $\epsilon''$ can be read from the (commutative) examples of real spectral triples defined by spin manifolds, where the $8$-periodicity of Clifford algebras is manifest. See \cite{Conn95a,Meye02a} for more details.

%%%%%%%%%%%%%%%%%%%%%%%%%%%%%%%%%%%%%%%%%%%%%%%%%%%%%%%%%%%%%%%%%%%%%%%%%%
\subsection{Gauge transformations and inner fluctuations}
%\label{}

Two spectral triples $(\algA, \ehH, \caD)$ and $(\algA', \ehH', \caD')$ are said to be unitary equivalent if there exists a unitary operator $U : \ehH \rightarrow \ehH'$ and an algebra isomorphism $\phi : \algA \rightarrow \algA'$ such that $\pi' \circ \phi = U \pi U^{-1}$, $\caD' = U \caD U^{-1}$, $J' = U J U^{-1}$, and $\gamma' = U \gamma U^{-1}$, when the operators $J$, $J'$, $\gamma$ and $\gamma'$ exist. 

A symmetry of a spectral triple is a unitary equivalence between two spectral triples such that $\ehH' = \ehH$, $\algA' = \algA$, and $\pi' = \pi$, so that $U : \ehH \rightarrow \ehH$ and $\phi \in \Aut(\algA)$. By definition, a symmetry acts only on $\caD$, $J$ and $\gamma$. Among these symmetries, we are only considering the automorphisms $\phi$ which are $\caA$-inner. This means that there is a unitary $u \in \caU(\caA)$ such that $\phi_u(a) = u a u^\invast$. This unitary in $\caA$ is sufficient to reconstruct all the symmetry, because then the unitary $U$ is given by $U = \pi(u) J \pi(u) J^{-1} : \ehH \rightarrow \ehH$. From the point of view of the bimodule structure on $\ehH$, $U$ is the conjugation with $\pi(u)$ since $\pi(u) J \pi(u) J^{-1} \Psi \simeq \pi(u) \Psi \pi(u)^\invast$. A straightforward computation shows that this inner symmetry leaves $J$ and $\gamma$ invariant. But the operator $\caD$ is modified, and one gets
\begin{equation*}
\caD^u = \caD + \pi(u) [\caD, \pi(u)^\invast] + \epsilon' J\left( \pi(u) [\caD, \pi(u)^\invast] \right) J^{-1}.
\end{equation*}
Interpreting a commutator with $\caD$ as a differential, this expression tells us that $\caD$ is modified by the addition of two inhomogeneous terms of the form ``$u \dd u^{-1}$''. Together, these two inhomogeneous terms produce a commutator or an anticommutator, depending on the sign of $\epsilon'$.

Inner symmetries of a spectral triple defines the gauge transformations in this setting. This is quite different to the general theory presented in \ref{subsec-Noncommutativeconnections} and the examples exposed in Section~\ref{sec-Derivation-basednoncommutativegeometry}. But we will see in a moment that in fact the two approaches can be reconciled.

In order to compensate for the inhomogeneous terms, we can use the same trick as in ordinary gauge field theory: add to the first order differential operator $\caD$ a gauge potential. 

In order to do that, we need to define the correct notion of noncommutative connections. The differential calculus is taken to be the universal differential calculus $(\Omega^\grast_U(\caA), \dd_U)$ and the right $\caA$-module is $\caA$. A noncommutative connection is then defined by a $1$-form $\omega = \sum_i a_i \dd_U b_i$ (finite sum). Elements in the vector spaces $\Omega^n_U(\caA)$ can be represented as bounded operators on $\ehH$:
\begin{equation*}
\pi_\caD\left(\sum_i a_i \dd_U b^1_i \cdots \dd_U b^n_i\right) = \sum_i \pi(a_i) [ \caD, \pi(b^1_i)] \cdots [ \caD, \pi(b^n_i)].
\end{equation*}
The map $\pi_\caD$ is not a representation of the graded algebra $\Omega^\grast_U(\caA)$, and $\dd_U$ is not represented by the commutator $[\caD, -]$ as a differential. Notice that $\pi_\caD$ can also be used to represent $n$-forms on the left module structure of $\ehH$ by the map $\sum_i a_i \dd_U b^1_i \cdots \dd_U b^n_i \mapsto J \pi_\caD\left(\sum_i a_i \dd_U b^1_i \cdots \dd_U b^n_i \right) J^{-1}$.

Using the bimodule structure on $\ehH$, one has the natural isomorphism $\ehH \simeq \caA \otimes_\caA \ehH \otimes_\caA \caA$, which identifies $\Psi \in \ehH$ with $\bbbone \otimes \Psi \otimes \bbbone$. Given a noncommutative connection $\widehat{\nabla} :  \caA \rightarrow \Omega^1_U(\caA)$, with $\omega = \widehat{\nabla} \bbbone$, we define the operator $\caD_\omega$ on $\ehH$ as:
\begin{equation*}
\caD_\omega (\Psi) = \pi_\caD(\omega) \Psi \otimes \bbbone + \bbbone \otimes \caD \Psi \otimes \bbbone + \epsilon' \bbbone \otimes \Psi \pi_\caD(\omega)^\invast,
\end{equation*}
for any $\Psi \in \ehH$. This operator can be written as $\caD_\omega = \caD + \pi_\caD(\omega) + \epsilon' J \pi_\caD(\omega) J^{-1}$.

There are now two ways to implement gauge transformations. The first one consists to look at it as an inner symmetry of the spectral triple. A direct computation shows that such an inner symmetry changes $\caD_\omega$ into
\begin{multline*}
(\caD_\omega)^u = \caD + \pi(u) \pi_\caD(\omega) \pi(u)^\invast + \pi(u) [\caD, \pi(u)^\invast]
\\
 + \epsilon' J \pi(u) \pi_\caD(\omega) \pi(u)^\invast J^{-1} + \epsilon' J \pi(u) [\caD, \pi(u)^\invast] J^{-1}.
\end{multline*}
The second way to implement a gauge transformation is to consider it as a gauge transformation on the right $\caA$-module $\caA$ by $a \mapsto u a$, as in \ref{subsec-Noncommutativeconnections}. This gauge transformation induces a gauge transformation on the bimodule $\caA \otimes_\caA \ehH \otimes_\caA \caA$, which is explicitly given by $\bbbone \otimes \Psi \otimes \bbbone \mapsto u \otimes \Psi \otimes u^\invast = u \Psi u^\invast = \pi(u) J \pi(u) J^{-1} \Psi$. The gauge transformation on $\widehat{\nabla}$ induces $\omega \mapsto \omega^u = u \omega u^\invast + u \dd_U u^\invast$, which in turn induces $\caD_\omega \mapsto \caD_{\omega^u}$ with
\begin{equation*}
\caD_{\omega^u} = \caD + \pi_\caD( u \omega u^\invast + u \dd_U u^\invast) + \epsilon' J \pi_\caD( u \omega u^\invast + u \dd_U u^\invast) J^{-1}.
\end{equation*}
This is exactly $(\caD_\omega)^u$. The two implementations of gauge transformations coincide.

It can be proved that if $(\algA, \ehH, \caD)$ is a spectral triple, then $(\algA, \ehH, \caD_\omega)$ is also a spectral triple. The replacement of the Dirac operator $\caD$ by $\caD_\omega$ is called an inner fluctuation in the space of Dirac operators associated to the couple $(\algA, \ehH)$. Some of the invariants defined by a spectral triple, for instance its class in $K$-homology, does not depend on these inner fluctuations.

In the case of a spectral triple associated to a spin geometry, an inner fluctuation looks very much like the twist of the Dirac operator by a connection defined on a vector bundle $\varE$. This procedure consists to replace $\slashed{S}$ by $\slashed{S} \otimes \varE$ and to define a new Dirac operator on this tensor product using a connection on $\varE$.

%%%%%%%%%%%%%%%%%%%%%%%%%%%%%%%%%%%%%%%%%%%%%%%%%%%%%%%%%%%%%%%%%%%%%%%%%%
\subsection{An elementary and instructive example}
\label{subsec-algebraCtwo}

Let us now consider the simple example of the algebra $\algA = \gC \oplus \gC$. This example has been used as a toy model for the Standard Model of particle physics \cite{Connes1990qp, MR1204452}, because it reveals a possible origin for the Higgs mechanism. Notice that this explanation for the Higgs mechanism is very similar to the one encountered in \ref{subsec-thematrixalgebra}.

The algebra $\algA = \gC \oplus \gC$ is commutative, and as such we can identify it with the space of (continuous, smooth) functions on a $2$ points space $\varM = \{ p_1, p_2 \}$. To any function $f \in C(\varM)$ we associate $f(p_1) \oplus f(p_2) \in \algA$. This identification is useful because the universal differential calculus of a commutative algebra of functions is easy to describe, as seen in \ref{subsec-Algebraicstructures}.

Here we have $\Omega^0_U(\gC \oplus \gC) = \gC \oplus \gC$. A $1$-form $f^0\dd_U f^1 \in \algA^{\otimes 2}$ defines only $2$ complex numbers $f^0(p_1) [f^1(p_2) - f^1(p_1)] = r_1$ and $f^0(p_2) [f^1(p_1) - f^1(p_2)] =  r_2$, so that $\Omega^1_U(\gC \oplus \gC) \simeq \gC \oplus \gC$. In the same way, a $2$-form takes only $2$ complex values, at $(p_1, p_2, p_1, p_2)$ and $(p_2, p_1, p_2, p_1)$.

The involution applied to the $1$-form $\omega = (r_1, r_2)$ gives $\omega^\invast = (- \overline{r_2}, - \overline{r_1})$. A connection $1$-form $\omega = (r_1, r_2)$ on the right module $\algA$ is compatible with the canonical Hermitian structure $\langle (z_1, z_2), (z'_1, z'_2) \rangle = (\overline{z_1} z'_1, \overline{z_2} z'_2 )$ if and only if $\overline{r_1} = r_2$. So that a  Hermitian connection is paramatrized by $\omega = (r, \overline{r})$. A straightforward computation of the curvature $\Omega = \dd_U \omega + \omega^2$ of this connection leads to $\Omega(p_1,p_2,p_1) = \Omega(p_2,p_1,p_2) = r + \overline{r} + r \overline{r}$. With $\phi = r + 1$, one has $r + \overline{r} + r \overline{r} = \overline{\phi}\phi -1$.

Consider now the spectral triple defined by the Hilbert space $\ehH = \gC^N \oplus \gC^N$ and the representation $\pi(z_1, z_2) (\Psi_1, \Psi_2) = (z_1 \Psi_1, z_2 \Psi_2)$ for any $(\Psi_1, \Psi_2) \in \ehH$. The Dirac operator is given by $\caD = \spmatrix{0 & M^\invast \\ M & 0}$, with $M \in M_N(\gC)$. This spectral triple is even with chirality $\gamma = \spmatrix{1 & 0 \\ 0 & -1}$. The representation of a $1$-form $\omega = (r_1, r_2)$ on $\ehH$ is given by $\pi_\caD(\omega) = \spmatrix{0 & r_1 M^\invast \\ r_2 M & 0}$ and the representation of the curvature $2$-form on $\ehH$ is given by
\begin{equation*}
\pi_\caD(\omega) = (\overline{\phi}\phi -1)\begin{pmatrix}
M^\invast M & 0  \\
0 & M M^\invast
\end{pmatrix}.
\end{equation*}
Then a natural Lagrangian for the dynamic of the gauge field is to consider the action
\begin{equation*}
S[\omega] = \tr (\pi_\caD(\omega)^2 ) = 2 (\overline{\phi}\phi -1)^2 \tr \left( (M^\invast M)^2 \right).
\end{equation*}
The fermionic part of the Lagrangian can be written as $S[\Psi, \omega] = \langle \Psi, \caD_\omega \Psi \rangle_{\ehH}$. The action $S[\omega]$ is zero for $\overline{\phi}\phi =1$, so that all the (non trivial) connections of the form $\omega = (e^{i\theta} -1, e^{-i\theta} -1)$ minimize the action. When looking at the fermionic part, these configurations contribute to mass terms for $\Psi \in \ehH$. This is a Higgs mechanism, for which the scalar fields come from the connection.

%%%%%%%%%%%%%%%%%%%%%%%%%%%%%%%%%%%%%%%%%%%%%%%%%%%%%%%%%%%%%%%%%%%%%%%%%%
\subsection{The Standard Model by Chamseddinne-Connes-Marcolli}
%\label{}

It is impossible to summarize in a few lines the construction of the noncommutative version of the Standard Model of particles based on the spectral triple approach. We will limit ourselves to describe the steps which allow to formalize it. The last version of this model is given in \cite{Chamseddine2006ep}. It is inspired by ideas exposed in \cite{Connes1996gi, Chamseddine1996zu}. Reviews and comments can be found in \cite{MR2179016, MR2371808, JureKrajSchu07a}. Previous versions of this models are described in \cite{Connes1990qp, MR1204452, Conn94, Martin1998363}.

Three mains steps are necessary to construct this model.

The first one concerns the general structure of the spectral triple. In its simplest form, gauge symmetries can be considered as symmetries implemented by functions on a space-time $\varM$ with values in a structure group $G$. In the spectral triple approach to gauge field theories, gauge symmetries are the inner automorphisms of an associative algebra. 

So far, following these constrains, in all the proposed models, the algebra is taken to be $C^\infty(\varM) \otimes \algA_F$, where $\algA_F$ is a finite dimensional algebra. This looks very much like the example presented in \ref{subsec-algebrafunctionsmatrices}, because any finite dimensional algebra is necessary a finite sum of matrix algebras.

To construct a spectral triple for such an algebra is facilitated by the following trick. Let $(\algA_1, \ehH_1, \caD_1)$ and $(\algA_2, \ehH_2, \caD_2,)$ be two even and real spectral triples, with chiralities $\gamma_1$ and $\gamma_2$ and realities $J_1$ and $J_2$. Then one can construct the even and real product spectral triple $(\algA, \ehH, \caD)$ with
\begin{align*}
\algA &= \algA_1 \otimes \algA_2,
&
\ehH &= \ehH_1 \otimes \ehH_2,
&
\caD &= \caD_1 \otimes 1 + \gamma_1 \otimes \caD_2,
&
\gamma &= \gamma_1 \otimes \gamma_2,
&
J &= J_1 \otimes J_2.
\end{align*}
The representation is $\pi(a_1 \otimes a_2)(\Psi_1 \otimes \Psi_2) = \pi_1(a_1)\Psi_1 \otimes \pi_2(a_2)\Psi_2$. The spectral triple of the Standard Model is constructed as the product of a commutative spectral triple $(C^\infty(\varM), L^2(\slashed{S}), \slashed{\partial})$ with a ``finite spectral triple'' $(\algA_F, \ehH_F, \caD_F)$. This defines what is called an almost commutative manifold.

The second step is to define a way to construct an action principle from a spectral triple. In \ref{subsec-algebraCtwo}, the action is defined as the trace of the square of the operator representing the curvature in the Hilbert space. A more subtle approach has been proposed in \cite{Chamseddine1996zu}. It is based on the spectral properties of the Dirac operator:
\begin{equation*}
S[\caD] = \tr \chi( \caD^2/\Lambda),
\end{equation*}
where $\tr$ is the trace on operators on $\ehH$, $\chi$ is a positive and even smooth function $\gR \rightarrow \gR$, and $\Lambda$ is a real (energy) cutoff which helps to make this trace well-behaved. For asymptotically large $\Lambda$, this action can be evaluated using heat kernel expansion.

This action has been evaluated for almost commutative geometries, and it gives rise to an action which contains at the same time the Einstein-Hilbert action and a Yang-Mills-Higgs action \cite{Chamseddine1996zu}. The coupling with fermions is taken to be $\langle \Psi, \caD \Psi \rangle_\ehH$ as before.

Finally, the last step is to find the correct finite spectral triple $(\algA_F, \ehH_F, \caD_F)$ in order to get close to the phenomenology of the usual Standard Model of particles physics. The algebra is the real algebra $\algA_F = \gC \oplus \gH \oplus M_3(\gC)$. The Hilbert space for one family of particles (and antiparticles) is $\ehH_F = M_4(\gC) \oplus M_4(\gC) \simeq \gC^{32}$. The full Hilbert space is $\ehH_F^{3}$ for the $3$ families. The representation of $\algA_F$ on $\ehH_F$ is given by left multiplication through the identification $\algA_F \subset M_4(\gC) \oplus M_4(\gC)$:
\begin{equation*}
\gC \oplus \gH \oplus M_3(\gC) \ni \lambda \oplus q \oplus m \mapsto
\begin{pmatrix}
\spmatrix{\lambda & 0 \\ 0 & \overline{\lambda}} & 0 \\
0 & q
\end{pmatrix}
\oplus
\begin{pmatrix}
\lambda & 0 \\
0 & m
\end{pmatrix}.
\end{equation*}
The Dirac operator $\caD_F$ is determined in terms of $3 \times 3$ Yukawa mixing matrices on $\ehH_F^{3}$. Left and right particles are in the respective $+1$ and $-1$ eigenspaces of the grading $\gamma_F$. Finally, the reality operator $J_F$ maps $\Psi_1 \oplus \Psi_2$ to $\Psi_2^\invast \oplus \Psi_1^\invast$.

As mentioned before, the spectral triple of the Standard Model is the product of a purely geometric spectral triple with this finite spectral triple. Inner fluctuations then give
\begin{equation*}
\caD_\omega = \slashed{\partial} + i \gamma^\mu A_\mu + \gamma^5 \caD_F + \gamma^5 \Phi,
\end{equation*}
where the $A_\mu$'s contain all the $U(1) \times SU(2) \times SU(3)$ gauge fields, and $\Phi$ is a doublet of scalar fields, which plays the role of Higgs fields. We refer to the references cited in the text for further details and comments.

As discuted at the end of \ref{subsec-algebrafunctionsmatrices}, almost commutative geometries correspond to trivial fiber bundles. In \cite{BoeiSuij11a}, spectral triples for non trivial situations are studied.

%%%%%%%%%%%%%%%%%%%%%%%%%%%%%%%%%%%%%%%%%%%%%%%%%%%%%%%%%%%%%%%%%%%%%%%%%%
\section{Conclusions}
%\label{}
%%%%%%%%%%%%%%%%%%%%%%%%%%%%%%%%%%%%%%%%%%%%%%%%%%%%%%%%%%%%%%%%%%%%%%%%%%

As presented in this review, noncommutative gauge field theories are multiform. For reasons of space, many examples were voluntarily omitted, for instance on the so famous noncommutative torus \cite{KrajWulk00a}. But the constructions of these theories rely on the ideas summarized here, and all their features are more or less similar to the one presented here.

Gauge field theories play an essential role in today physics. Unfortunately, some problems they generate need further investigations, particularly concerning quantization (gauge fixing, BRS symmetries) and the adjonction of scalar fields to implement a Higgs mechanism. We saw how the second problem gets an elegant solution in the framework of noncommutative gauge field theories. As a ``proof of concept'', the Standard Model of particle physics can be written in this language.

As seen in this review, the mathematical structure which permits to construct a good gauge field theory is based on some elementary ingredients:
\begin{itemize}
\item A general structure which encodes the (local) symmetries: in ordinary geometry, this is the principal fiber bundle, in noncommutative geometry, this is the associative algebra.

\item A correct theory of representations of these symmetries in which to place matter fields: in ordinary geometry, this is the theory of associated vector bundles, in noncommutative geometry, this is the module structures.

\item A differential structure to encode the differential aspect of the theory (covariant derivatives, connections $1$-forms…): in ordinary geometry, the de~Rham calculus (with values in a Lie algebra) is used, in noncommutative geometry, the main concept is a differential calculus in general, but implemented in possibly different ways.
\end{itemize}
Ordinary geometry and noncommutative geometry are not the only theories with these characteristics. For instance, the theory of transitive Lie algebroids \cite{Mack05a} leads also to a natural and to a rich theory of connections which share a lot in common with some of the theories of connections defined in noncommutative geometry \cite{Mass38,Mass41}.

We hope that these new mathematical horizons will stimulate new ideas in the phenomenology of particle physics. Because these theories go beyond the so powerful theory of group representations used until now, we can expect that some of the challenging problems ``beyond the Standard Model'' will finally find a satisfactory solution in these richer structures.

\bibliography{bibliographie}

\begin{thebibliography}{79}
\providecommand{\natexlab}[1]{#1}
\providecommand{\url}[1]{\texttt{#1}}
\expandafter\ifx\csname urlstyle\endcsname\relax
  \providecommand{\doi}[1]{doi: #1}\else
  \providecommand{\doi}{doi: \begingroup \urlstyle{rm}\Url}\fi

\bibitem[Atiyah(1966)]{Atiy66a}
M.~F. Atiyah.
\newblock ${K}$-theory and reality.
\newblock \emph{Quart. J. Math. Oxford}, 17:\penalty0 367--386, 1966.

\bibitem[Bertlmann(1996)]{Bert96}
R.~Bertlmann.
\newblock \emph{Anomalies in Quantum Field Theory}.
\newblock Oxford Science Publications, 1996.

\bibitem[Blackadar(1998)]{Blac98}
B.~Blackadar.
\newblock \emph{{$K$}-Theory for Operator Algebras}, volume~5 of \emph{Math.
  Sc. Research Inst. Pub.}
\newblock Cambridge University Press, 1998.

\bibitem[Blackadar(2006)]{Blac06}
B.~Blackadar.
\newblock \emph{Operator Algebras, Theory of {$C^\ast$}-Algebras and von
  {N}eumann Algebras}, volume 122 of \emph{Encyclopaedia of Mathematical
  Sciences}.
\newblock Springer-Verlag, 2006.

\bibitem[Boeijink and van Suijlekom(2011)]{BoeiSuij11a}
J.~Boeijink and W.~D. van Suijlekom.
\newblock The noncommutative geometry of {Y}ang-{M}ills fields.
\newblock \emph{J.Geom.Phys.}, 61:\penalty0 1122--1134, 2011.
\newblock URL \url{http://arxiv.org/abs/1008.5101}.

\bibitem[Bratteli and Robinson(2002--2003)]{BratRobi02c}
O.~Bratteli and D.~W. Robinson.
\newblock \emph{Operator Algebras and Quantum Statistical Mechanics 1 \& 2}.
\newblock Theoretical and Mathematical Physics. Springer-Verlag, 2002--2003.

\bibitem[Cagnache et~al.(2011)Cagnache, Masson, and Wallet]{Mass32}
E.~Cagnache, T.~Masson, and J.-C. Wallet.
\newblock Noncommutative {Y}ang-{M}ills-{H}iggs actions from derivation-based
  differential calculus.
\newblock \emph{Journal of Noncommutative Geometry}, 5\penalty0 (1):\penalty0
  39--67, 2011.
\newblock URL \url{http://arxiv.org/abs/0804.3061}.
\newblock arXiv:0804.3061, High Energy Physics - Theory (hep-th).

\bibitem[Cartan and Eilenberg(1956)]{CartEile56}
H.~Cartan and S.~Eilenberg.
\newblock \emph{Homological Algebra}.
\newblock Princeton University Press, 1956.

\bibitem[Chamseddine and Connes(1997)]{Chamseddine1996zu}
A.~H. Chamseddine and A.~Connes.
\newblock {The spectral action principle}.
\newblock \emph{Commun. Math. Phys.}, 186:\penalty0 731--750, 1997.
\newblock \doi{10.1007/s002200050126}.

\bibitem[Chamseddine et~al.(2007)Chamseddine, Connes, and
  Marcolli]{Chamseddine2006ep}
A.~H. Chamseddine, A.~Connes, and M.~Marcolli.
\newblock {Gravity and the standard model with neutrino mixing}.
\newblock \emph{Adv. Theor. Math. Phys.}, 11:\penalty0 991--1089, 2007.

\bibitem[Chari and Pressley(1994)]{CharPres94}
V.~Chari and A.~Pressley.
\newblock \emph{A Guide to Quantum Groups}.
\newblock Cambridge University Press, 1994.

\bibitem[Connes(1985)]{Conn85}
A.~Connes.
\newblock \emph{Non-commutative differential geometry}.
\newblock Publications Math{\'e}matiques de l'{IHES}, 1985.

\bibitem[Connes(1994)]{Conn94}
A.~Connes.
\newblock \emph{Noncommutative Geometry}.
\newblock Academic Press, 1994.

\bibitem[Connes(1995)]{Conn95a}
A.~Connes.
\newblock Noncommutative geometry and reality.
\newblock \emph{J. Math. Phys.}, 36\penalty0 (11):\penalty0 6194--6231, 1995.

\bibitem[Connes(1996)]{Connes1996gi}
A.~Connes.
\newblock {Gravity coupled with matter and the foundation of non- commutative
  geometry}.
\newblock \emph{Commun. Math. Phys.}, 182:\penalty0 155--176, 1996.
\newblock \doi{10.1007/BF02506388}.

\bibitem[Connes(2000)]{Connes2000by}
A.~Connes.
\newblock {A short survey of noncommutative geometry}.
\newblock \emph{Journal of Mathematical Physics}, 41:\penalty0 3832--3866,
  2000.
\newblock \doi{10.1063/1.533329}.

\bibitem[Connes and Lott(1990)]{Connes1990qp}
A.~Connes and J.~Lott.
\newblock Particle models and noncommutative geometry.
\newblock \emph{Nuclear Phys. B Proc. Suppl.}, 18B:\penalty0 29--47 (1991),
  1990.
\newblock ISSN 0920-5632.
\newblock \doi{10.1016/0920-5632(91)90120-4}.
\newblock URL \url{http://dx.doi.org/10.1016/0920-5632(91)90120-4}.
\newblock Recent advances in field theory (Annecy-le-Vieux, 1990).

\bibitem[Connes and Lott(1992)]{MR1204452}
A.~Connes and J.~Lott.
\newblock The metric aspect of noncommutative geometry.
\newblock In \emph{New symmetry principles in quantum field theory
  ({C}arg{\`e}se, 1991)}, volume 295 of \emph{NATO Adv. Sci. Inst. Ser. B
  Phys.}, pages 53--93. Plenum, New York, 1992.

\bibitem[Connes and Marcolli(2008{\natexlab{a}})]{ConnMarc08a}
A.~Connes and M.~Marcolli.
\newblock A walk in the noncommutative garden.
\newblock In M.~Khalkhali and M.~Marcolli, editors, \emph{An invitation to
  noncommutative geometry}, pages 1--128. World Scientific Publishing Company,
  2008{\natexlab{a}}.

\bibitem[Connes and Marcolli(2008{\natexlab{b}})]{MR2371808}
A.~Connes and M.~Marcolli.
\newblock \emph{Noncommutative geometry, quantum fields and motives}, volume~55
  of \emph{American Mathematical Society Colloquium Publications}.
\newblock American Mathematical Society, Providence, RI, 2008{\natexlab{b}}.
\newblock ISBN 978-0-8218-4210-2.

\bibitem[Coquereaux(1992{\natexlab{a}})]{MR1150534}
R.~Coquereaux.
\newblock Connections, metrics, symmetries and scalar fields in commutative and
  noncommutative geometry.
\newblock \emph{Classical Quantum Gravity}, 9\penalty0 (suppl.):\penalty0
  S41--S53, 1992{\natexlab{a}}.
\newblock ISSN 0264-9381.
\newblock URL \url{http://stacks.iop.org/0264-9381/9/41}.
\newblock Les Journ{{\'e}}es Relativistes (Carg{{\`e}}se, 1991).

\bibitem[Coquereaux(1992{\natexlab{b}})]{MR1205605}
R.~Coquereaux.
\newblock Yang-{M}ills fields and symmetry breaking: from {L}ie super-algebras
  to noncommutative geometry.
\newblock In \emph{Quantum groups and related topics ({W}roc\l aw, 1991)},
  volume~13 of \emph{Math. Phys. Stud.}, pages 115--127. Kluwer Acad. Publ.,
  Dordrecht, 1992{\natexlab{b}}.

\bibitem[Coquereaux et~al.(1991)Coquereaux, Esposito-Farese, and
  Vaillant]{MR1103056}
R.~Coquereaux, G.~Esposito-Farese, and G.~Vaillant.
\newblock Higgs field as {Y}ang-{M}ills fields and discrete symmetries.
\newblock \emph{Nuclear Phys. B}, 353\penalty0 (3):\penalty0 689--706, 1991.
\newblock ISSN 0550-3213.
\newblock \doi{10.1016/0550-3213(91)90323-P}.
\newblock URL \url{http://dx.doi.org/10.1016/0550-3213(91)90323-P}.

\bibitem[Cuntz and Khalkhali(1997)]{CuntKhal97}
J.~Cuntz and M.~Khalkhali, editors.
\newblock \emph{Cyclic Cohomology and Noncommutative Geometry}, volume~17 of
  \emph{Fields Institute Communications}.
\newblock AMS, 1997.

\bibitem[de~Goursac et~al.(2007)de~Goursac, Wallet, and
  Wulkenhaar]{DEGOURSAC2007HAL-001359171}
A.~de~Goursac, J.-C. Wallet, and R.~Wulkenhaar.
\newblock {N}oncommutative {I}nduced {G}auge {T}heory.
\newblock \emph{European Physical Journal C}, 51:\penalty0 977--987, 2007.
\newblock URL \url{http://hal.archives-ouvertes.fr/hal-00135917/en/}.

\bibitem[de~Goursac et~al.(2008)de~Goursac, Wallet, and
  Wulkenhaar]{deGoursac2008rb}
A.~de~Goursac, J.-C. Wallet, and R.~Wulkenhaar.
\newblock {On the vacuum states for noncommutative gauge theory}.
\newblock \emph{European Physical Journal C}, 56:\penalty0 293--304, 2008.

\bibitem[Dubois-Violette(1988)]{DuVi88}
M.~Dubois-Violette.
\newblock D{\'e}rivations et calcul differentiel non commutatif.
\newblock \emph{C.R. Acad. Sci. Paris, S{\'e}rie {I}}, 307:\penalty0 403--408,
  1988.

\bibitem[Dubois-Violette(1991)]{DuVi91}
M.~Dubois-Violette.
\newblock Non-commutative differential geometry, quantum mechanics and gauge
  theory.
\newblock In C.~Bartocci, U.~Bruzzo, and R.~Cianci, editors, \emph{Differential
  Geometric Methods in Theoretical Physics, Proceedings of the 19th
  International Conference Held in Rapallo, Italy 19--24 June 1990}, volume 375
  of \emph{Lecture Notes in Physics}. Springer-Verlag, 1991.

\bibitem[Dubois-Violette(1997)]{MR1443921}
M.~Dubois-Violette.
\newblock Some aspects of noncommutative differential geometry.
\newblock In \emph{Geometry and nature ({M}adeira, 1995)}, volume 203 of
  \emph{Contemp. Math.}, pages 145--157. Amer. Math. Soc., Providence, RI,
  1997.

\bibitem[Dubois-Violette(2001)]{DuVi01a}
M.~Dubois-Violette.
\newblock Lectures on graded differential algebras and noncommutative geometry.
\newblock In Y.~Maeda and H.~Moriyoshi, editors, \emph{{Noncommutative
  Differential Geometry and Its Applications to Physics}}. Kluwer Academic
  Publishers, 2001.

\bibitem[Dubois-Violette and Masson(1996)]{Mass08}
M.~Dubois-Violette and T.~Masson.
\newblock On the first order operators in bimodules.
\newblock \emph{Letters in Mathematical Physics}, 37:\penalty0 467--474, 1996.
\newblock URL \url{http://arxiv.org/abs/q-alg/9507028}.

\bibitem[Dubois-Violette and Masson(1998)]{Mass14}
M.~Dubois-Violette and T.~Masson.
\newblock ${SU}(n)$-connections and noncommutative differential geometry.
\newblock \emph{Journal of Geometry and Physics}, 25\penalty0 (1,2):\penalty0
  104, 1998.
\newblock URL \url{http://arxiv.org/abs/dg-ga/9612017}.

\bibitem[Dubois-Violette and Michor(1994)]{DuViMich94}
M.~Dubois-Violette and P.~W. Michor.
\newblock D{\'e}rivations et calcul differentiel non commutatif {II}.
\newblock \emph{C.R. Acad. Sci. Paris, S{\'e}rie {I}}, 319:\penalty0 927--931,
  1994.

\bibitem[Dubois-Violette and Michor(1996)]{DuViMich96}
M.~Dubois-Violette and P.~W. Michor.
\newblock Connections on central bimodules in noncommutative differential
  geometry.
\newblock \emph{Journal of Geometry and Physics}, 20:\penalty0 218--232, 1996.

\bibitem[Dubois-Violette and Michor(1997)]{DuViMich97}
M.~Dubois-Violette and P.~W. Michor.
\newblock More on the {F}r{\"o}licher-{N}ijenhuis bracket in non commutative
  differential geometry.
\newblock \emph{Journal of Pure and Applied Algebra}, 121:\penalty0 107--135,
  1997.

\bibitem[Dubois-Violette et~al.(1990{\natexlab{a}})Dubois-Violette, Kerner, and
  Madore]{DuViKernMado90a}
M.~Dubois-Violette, R.~Kerner, and J.~Madore.
\newblock Noncommutative differential geometry of matrix algebras.
\newblock \emph{Journal of Mathematical Physics}, 31:\penalty0 316,
  1990{\natexlab{a}}.

\bibitem[Dubois-Violette et~al.(1990{\natexlab{b}})Dubois-Violette, Kerner, and
  Madore]{DuViKernMado90b}
M.~Dubois-Violette, R.~Kerner, and J.~Madore.
\newblock Noncommutative differential geometry and new models of gauge theory.
\newblock \emph{Journal of Mathematical Physics}, 31:\penalty0 323,
  1990{\natexlab{b}}.

\bibitem[Fournel et~al.(2011)Fournel, Lazzarini, and Masson]{Mass41}
C.~Fournel, S.~Lazzarini, and T.~Masson.
\newblock Local description of generalized forms on transitive {L}ie algebroids
  and applications.
\newblock arxiv 1109.4282v1, 2011.
\newblock URL \url{http://arxiv.org/abs/1109.4282v1}.

\bibitem[Gayral et~al.(2004)Gayral, Gracia-Bond{\'\i}a, Iochum, Sch{\"u}cker,
  and V{\'a}rilly]{Gayral2003dm}
V.~Gayral, J.~M. Gracia-Bond{\'\i}a, B.~Iochum, T.~Sch{\"u}cker, and J.~C.
  V{\'a}rilly.
\newblock {Moyal planes are spectral triples}.
\newblock \emph{Commun. Math. Phys.}, 246:\penalty0 569--623, 2004.
\newblock \doi{10.1007/s00220-004-1057-z}.

\bibitem[Gracia-Bond{\'\i}a and
  V{\'a}rilly(1988{\natexlab{a}})]{GraciaBondia1987kw}
J.~M. Gracia-Bond{\'\i}a and J.~C. V{\'a}rilly.
\newblock {Algebras of distributions suitable for phase space quantum
  mechanics. I}.
\newblock \emph{Journal of Mathematical Physics}, 29:\penalty0 869--879,
  1988{\natexlab{a}}.
\newblock \doi{10.1063/1.528200}.

\bibitem[Gracia-Bond{\'\i}a and V{\'a}rilly(1988{\natexlab{b}})]{Varilly1988jk}
J.~M. Gracia-Bond{\'\i}a and J.~C. V{\'a}rilly.
\newblock {Algebras of distributions suitable for phase-space quantum
  mechanics. {II}. Topologies on the Moyal algebra}.
\newblock \emph{Journal of Mathematical Physics}, 29:\penalty0 880--887,
  1988{\natexlab{b}}.
\newblock \doi{10.1063/1.527984}.

\bibitem[Gracia-Bond{\'\i}a et~al.(2001)Gracia-Bond{\'\i}a, V{\'a}rilly, and
  Figueroa]{GracVariFigu01}
J.~M. Gracia-Bond{\'\i}a, J.~C. V{\'a}rilly, and H.~Figueroa.
\newblock \emph{Elements of Noncommutative Geometry}.
\newblock Birkha{\"u}ser Advanced Texts. Birkha{\"u}ser, 2001.

\bibitem[Grosse and Wohlgenannt(2006)]{Grosse2006hh}
H.~Grosse and M.~Wohlgenannt.
\newblock {Noncommutative QFT and renormalization}.
\newblock \emph{Journal of Physics: Conference Series}, 53:\penalty0 764--792,
  2006.
\newblock \doi{10.1088/1742-6596/53/1/050}.

\bibitem[Grosse and Wohlgenannt(2007)]{Grosse2007dm}
H.~Grosse and M.~Wohlgenannt.
\newblock {Induced Gauge Theory on a Noncommutative Space}.
\newblock \emph{European Physical Journal C}, 52:\penalty0 435--450, 2007.
\newblock \doi{10.1140/epjc/s10052-007-0369-5}.

\bibitem[Grosse and Wulkenhaar(2005{\natexlab{a}})]{GrosWulk05a}
H.~Grosse and R.~Wulkenhaar.
\newblock Renormalisation of $\phi^4$-theory on noncommutative $\mathbb{R}^4$
  in the matrix base.
\newblock \emph{Comm. Math. Phys.}, 256:\penalty0 305, 2005{\natexlab{a}}.

\bibitem[Grosse and Wulkenhaar(2005{\natexlab{b}})]{GrosWulk05b}
H.~Grosse and R.~Wulkenhaar.
\newblock Power-counting theorem for non-local matrix models and
  renormalisation.
\newblock \emph{Comm. Math. Phys.}, 254:\penalty0 91, 2005{\natexlab{b}}.

\bibitem[Higson and Roe(2004)]{HigsRoe04}
N.~Higson and J.~Roe.
\newblock \emph{Analytic {$K$}-Homology}.
\newblock Oxford Science Publications, 2004.

\bibitem[Jacobson(1985)]{Jaco85}
N.~Jacobson.
\newblock \emph{Basic Algebra, vol. {I} \& {II}}.
\newblock Freeman, 1985.

\bibitem[Jureit et~al.(2007)Jureit, Krajewski, Sch{\"u}cker, and
  Stephan]{JureKrajSchu07a}
J.-H. Jureit, T.~Krajewski, T.~Sch{\"u}cker, and C.~A. Stephan.
\newblock On the noncommutative standard model.
\newblock \emph{Acta Phys. Polon. B}, 38\penalty0 (10):\penalty0 3181--3202,
  2007.

\bibitem[Kadison and Ringrose(1997)]{KadiRing97c}
R.~Kadison and J.~Ringrose.
\newblock \emph{Fundamentals of the Theory of Operator Algebras, volumes {I},
  {II}}.
\newblock Graduate Studies in Mathematics. AMS, 1997.

\bibitem[Klimyk and Schmudgen(1997)]{KlimSchm97}
A.~Klimyk and K.~Schmudgen.
\newblock \emph{Quantum Groups and Their Representations}.
\newblock TMP. Springer-Verlag, 1997.

\bibitem[Kobayashi and Nomizu(1996)]{KobaNomi96c}
S.~Kobayashi and K.~Nomizu.
\newblock \emph{Foundations of Differential Geometry, vol. 1 \& 2}.
\newblock Wiley Classics Library. Interscience Publishers, 1996.

\bibitem[Krajewski and Wulkenhaar(2000)]{KrajWulk00a}
T.~Krajewski and R.~Wulkenhaar.
\newblock Perturbative quantum gauge fields on the noncommutative torus.
\newblock \emph{Int. J. Mod. Phys.}, A15:\penalty0 1011, 2000.

\bibitem[Landi(1997)]{Land97a}
G.~Landi.
\newblock \emph{{An Introduction to Noncommutative Spaces and Their
  Geometries}}.
\newblock Springer-Verlag, 1997.

\bibitem[Lazzarini and Masson(2012)]{Mass38}
S.~Lazzarini and T.~Masson.
\newblock Connections on {L}ie algebroids and on derivation-based
  non-commutative geometry.
\newblock \emph{Journal of Geometry and Physics (to appear)}, 2012.
\newblock URL \url{http://arxiv.org/abs/1003.6106}.
\newblock arxiv 1003.6106.

\bibitem[Loday(1998)]{Loda98}
J.-L. Loday.
\newblock \emph{Cyclic Homology}, volume 301 of \emph{Grundlehren der
  mathematischen Wissenschaften}.
\newblock Springer-Verlag, 2 edition, 1998.

\bibitem[Mackenzie(2005)]{Mack05a}
K.~Mackenzie.
\newblock \emph{General Theory of {L}ie Groupoids and {L}ie Algebroids}.
\newblock Number 213 in London Mathematical Society Lecture Note Series.
  Cambridge University Press, 2005.

\bibitem[Mart{\'\i}n et~al.(1998)Mart{\'\i}n, Gracia-Bond{\'\i}a, and
  V{\'a}rilly]{Martin1998363}
C.~P. Mart{\'\i}n, J.~M. Gracia-Bond{\'\i}a, and J.~C. V{\'a}rilly.
\newblock The standard model as a noncommutative geometry: the low-energy
  regime.
\newblock \emph{Physics Reports}, 294\penalty0 (6):\penalty0 363 -- 406, 1998.
\newblock ISSN 0370-1573.
\newblock \doi{10.1016/S0370-1573(97)00053-7}.
\newblock URL
  \url{http://www.sciencedirect.com/science/article/pii/S0370157397000537}.

\bibitem[Masson(1995)]{Mass11}
T.~Masson.
\newblock \emph{G{\'e}om{\'e}trie non commutative et applications {\`a} la
  th{\'e}orie des champs}.
\newblock Th{\`e}se de doctorat, Universit{\'e} Paris XI, 1995.
\newblock Th{\`e}se soutenue le 13 d{\'e}cembre 1995.

\bibitem[Masson(1996)]{Mass07}
T.~Masson.
\newblock Submanifolds and quotient manifolds in noncommutative geometry.
\newblock \emph{Journal of Mathematical Physics}, 37\penalty0 (5):\penalty0
  2484--2497, 1996.
\newblock URL \url{http://arxiv.org/abs/q-alg/9507030}.

\bibitem[Masson(1999)]{Mass15}
T.~Masson.
\newblock On the noncommutative geometry of the endomorphism algebra of a
  vector bundle.
\newblock \emph{Journal of Geometry and Physics}, 31:\penalty0 142, 1999.
\newblock URL \url{http://arxiv.org/abs/math.DG/9803088}.

\bibitem[Masson(2008{\natexlab{a}})]{Mass28}
T.~Masson.
\newblock An informal introduction to the ideas and concepts of noncommutative
  geometry.
\newblock In F.~H{\'e}lein and J.~Kouneiher, editors, \emph{{I}ntegrables
  {S}ystems and {Q}uantum {F}ields {T}heories}, Physique-Math{\'e}matique,
  Travaux en cours. {H}ermann, 2008{\natexlab{a}}.
\newblock URL \url{http://arxiv.org/abs/math-ph/0612012}.
\newblock Lecture given at the 6th Peyresq meeting 'Integrable systems and
  quantum field theory'.

\bibitem[Masson(2008{\natexlab{b}})]{Mass30}
T.~Masson.
\newblock Noncommutative generalization of ${SU}(n)$-principal fiber bundles: a
  review.
\newblock \emph{Journal of Physics: Conference Series}, 103:\penalty0 012003,
  2008{\natexlab{b}}.
\newblock URL
  \url{http://iopscience.iop.org/1742-6596/103/1/012003/pdf/1742-6596_103_1_012003.pdf}.

\bibitem[Masson(2008{\natexlab{c}})]{Mass31}
T.~Masson.
\newblock \emph{Introduction aux (Co)Homologies}.
\newblock Physique-Math{\'e}matique, Travaux en cours. Hermann,
  2008{\natexlab{c}}.

\bibitem[Masson(2008{\natexlab{d}})]{Mass33}
T.~Masson.
\newblock Examples of derivation-based differential calculi related to
  noncommutative gauge theories.
\newblock \emph{International Journal of Geometric Methods in Modern Physics},
  5\penalty0 (8):\penalty0 1315--1336, 2008{\natexlab{d}}.
\newblock URL \url{http://arxiv.org/abs/0810.4815}.

\bibitem[Masson and S{\'e}ri{\'e}(2005)]{Mass25}
T.~Masson and E.~S{\'e}ri{\'e}.
\newblock Invariant noncommutative connections.
\newblock \emph{Journal of Mathematical Physics}, 46:\penalty0 123503, 2005.
\newblock URL \url{http://arxiv.org/abs/math-ph/0407022}.

\bibitem[Meyer(2002)]{Meye02a}
R.~Meyer.
\newblock \emph{Real Spectral Triples and Charge Conjugation}, pages 11--20.
\newblock Volume 596 of  \citet{ScheHaraWern02a}, 2002.
\newblock \doi{10.1007/3-540-46082-9_2}.

\bibitem[O'Raifeartaigh and Straumann(2000)]{o2000gauge}
L.~O'Raifeartaigh and N.~Straumann.
\newblock Gauge theory: Historical origins and some modern developments.
\newblock \emph{Reviews of Modern Physics}, 72\penalty0 (1):\penalty0 1, 2000.

\bibitem[Peskin and Schroeder(2008)]{PeskSchr08a}
M.~E. Peskin and D.~V. Schroeder.
\newblock \emph{An Introduction to Quantum Field Theory}.
\newblock Perseus Books, 2008.

\bibitem[{R}ivasseau(2007)]{RIVASSEAU2007HAL-001656861}
V.~{R}ivasseau.
\newblock {N}on-commutative {R}enormalization.
\newblock In \emph{Quantum Spaces}, volume~53 of \emph{Progress in Mathematical
  Physics}. Birkha{\"u}ser, 2007.
\newblock URL \url{http://hal.archives-ouvertes.fr/hal-00165686/en/}.
\newblock 82 pages, {P}oincare {S}eminar, updated and expanded version of
  ar{X}iv:hep-th/0702068 07-63.

\bibitem[R{\o}rdam et~al.(2000)R{\o}rdam, Larsen, and Laustsen]{RordLarsLaus00}
M.~R{\o}rdam, F.~Larsen, and N.~J. Laustsen.
\newblock \emph{An Introduction to {$K$}-theory for {$C^\ast$}-Algebras}.
\newblock Cambridge University Press, 2000.

\bibitem[Scheck et~al.(2002)Scheck, Upmeier, and Werner]{ScheHaraWern02a}
F.~Scheck, H.~Upmeier, and W.~Werner, editors.
\newblock \emph{Noncommutative Geometry and the Standard Model of Elementary
  Particle Physics}, volume 596 of \emph{Lecture Notes in Physics}.
\newblock Springer-Verlag, 2002.

\bibitem[Sch{\"u}cker(2005)]{MR2179016}
T.~Sch{\"u}cker.
\newblock Forces from {C}onnes' geometry.
\newblock In \emph{Topology and geometry in physics}, volume 659 of
  \emph{Lecture Notes in Phys.}, pages 285--350. Springer, Berlin, 2005.

\bibitem[Serre(1955)]{Serr55a}
J.-P. Serre.
\newblock Faisceaux alg{\'e}briques coh{\'e}rents.
\newblock \emph{Ann. of Math. (2)}, 61:\penalty0 197--278, 1955.

\bibitem[Swan(1962)]{Swan62a}
R.~G. Swan.
\newblock Vector bundles and projective modules.
\newblock \emph{Trans. Amer. Math. Soc.}, 105:\penalty0 264--277, 1962.

\bibitem[Takesaki(2002, 2003)]{Take02c}
M.~Takesaki.
\newblock \emph{Theory of Operator Algebras {I}, {II}, {III}}.
\newblock Encyclopaedia of Mathematical Sciences. Springer-Verlag, 2002, 2003.

\bibitem[Wallet(2008)]{WALLET2007HAL-001709651}
J.-C. Wallet.
\newblock {N}oncommutative {I}nduced {G}auge {T}heories on {M}oyal {S}paces.
\newblock \emph{Journal of Physics: Conference Series}, 103:\penalty0 012207,
  2008.
\newblock URL \url{http://hal.archives-ouvertes.fr/hal-00170965/en/}.
\newblock 24 pages, 6 figures. {T}alk given at the ``{I}nternational
  {C}onference on {N}oncommutative {G}eometry and {P}hysics'', {A}pril 2007,
  {O}rsay ({F}rance) 07-60.

\bibitem[Wegge-Olsen(1993)]{Wegg93}
N.~E. Wegge-Olsen.
\newblock \emph{{$K$}-Theory and {$C^\ast$}-Algebras}.
\newblock Oxford University Press, 1993.

\bibitem[Weibel(1997)]{Weib97}
C.~A. Weibel.
\newblock \emph{An Introduction to Homological Algebra}, volume~38 of
  \emph{Cambridge studies in advanced mathematics}.
\newblock Cambridge University Press, 1997.

\end{thebibliography}

\end{document}